\documentclass{aa}

\PassOptionsToPackage{update,prepend=false}{epstopdf}
\usepackage[varg]{txfonts}

\usepackage{lineno}
\usepackage{natbib}
\usepackage{amsmath}
\usepackage{graphicx}
\usepackage{xspace}
\usepackage{xcolor}
\usepackage[markup=underlined]{changes}

\usepackage{hyperref}
\hypersetup{
    colorlinks=true,
    linkcolor={red!50!black},
    citecolor={blue!70!black},
    urlcolor={blue!80!black}
}

\usepackage[normalem]{ulem}
\usepackage{multirow}

\usepackage{etoolbox}
\newtoggle{thesis}

\begin{document}

\title{Black-hole X-ray binary Swift J1727.8$-$1613 shows simultaneous Type-B and Type-C quasi-periodic oscillations across the hard-intermediate and soft-intermediate states}
\titlerunning{Type-B QPO spanning the hard-intermediate and soft-intermediate states}

\author{Pei~Jin\inst{1}\thanks{peijin@astro.rug.nl} \and
Mariano~M\'{e}ndez\inst{1}\thanks{mariano@astro.rug.nl} \and
Federico~Garc\'{\i}a\inst{2} \and
Diego~Altamirano\inst{3} \and
Federico M. Vincentelli\inst{4,3}}

\institute{Kapteyn Astronomical Institute, University of Groningen, P.O.\ BOX 800, 9700 AV Groningen, The Netherlands
\and Instituto Argentino de Radioastronom\'{\i}a (CCT La Plata, CONICET; CICPBA; UNLP), C.C.5, (1894) Villa Elisa, Buenos Aires, Argentina
\and School of Physics and Astronomy, University of Southampton, Southampton, Hampshire SO17 1BJ, UK
\and INAF Istituto di Astrofisica e Planetologia Spaziali, Via del Fosso del Cavaliere 100, 00133 Roma, Italy
}

\date{Received xxx; accepted xxx}

\abstract{

We present a timing analysis of \textit{Insight}-HXMT observations of the black-hole X-ray binary Swift J1727.8$-$1613 across a bright soft X-ray flare on 2023 September 19 (MJD 60206). At the peak of the flare, the source undergoes a brief transition from the hard-intermediate state (HIMS) into the soft-intermediate state (SIMS), marked by the simultaneous appearance of three discrete radio jet ejections, a drop in broadband noise in the 2$-$10 keV band, and the presence of a narrow quasi-periodic oscillation (QPO) with a characteristic ``U''-shaped phase-lag spectrum and a quality factor of $Q \geq 6$, features that robustly identify it as a Type-B QPO. The Type-C QPO, which was clearly detected in the HIMS prior to the flare, is not observed at the flare's peak and only reappears afterward. Most notably, we find that the Type-B QPO is not restricted to the SIMS: it is present throughout all our observations, including those taken in the HIMS, where it appears as a broad shoulder of the Type-C QPO. During the flare, the Type-B and Type-C QPOs exhibit distinct evolutionary trends in frequency, fractional rms amplitude, and phase lag. These results challenge the traditional view that Type-B QPOs are exclusive to the SIMS, a state that is, in fact, defined by their appearance in the power spectrum, and directly linked to discrete jet ejections. Instead, our findings suggest that the physical conditions giving rise to Type-B QPOs occur more broadly within the inner accretion flow.

}

\keywords{accretion, accretion discs -- stars: individual: Swift~J1727.8–1613 -- stars: black holes -- X-rays: binaries
}

\maketitle 



\section{Introduction}
\label{sec:introduction}

Black-hole X-ray binaries (BHXBs) are typically transient systems that stay most of the time in the quiescent state, but show recurrent bright X-ray outbursts lasting for weeks to months~\citep[e.g.][]{1996ARA&A..34..607T, 2006csxs.book..157M}. During the outbursts, the time-averaged X-ray energy spectrum typically consists of a thermal and a power-law (PL) component~\citep[e.g.][]{2006csxs.book..157M}. The thermal emission originates from the accretion disc~\citep{1973A&A....24..337S} and is well described by a multi-temperature black-body with a characteristic temperature around 1.0 keV. The PL component is attributed to the so-called corona with temperatures of up to $\sim$100~keV~\citep[e.g.][]{1980A&A....86..121S}, in which soft photons from the accretion disc are inverse-Compton scattered. 

Short time-scale variability ranging from milliseconds to hundreds of seconds is very common in BHXBs~\citep[see][and references therein]{2011BASI...39..409B, 2019NewAR..8501524I}. Quasi-periodic oscillations (QPOs), appearing as narrow peaks in the power spectrum, are a prominent feature in BHXBs. Low-frequency QPOs (ranging from 0.1 to 30 Hz) are typically classified into three subtypes—Type A, Type B, and Type C—based on their centroid frequency, amplitude, quality factor, and the characteristics of the accompanying broadband noise~\citep{1999ApJ...526L..33W, 2002ApJ...564..962R, 2005ApJ...629..403C}.

Based on the X-ray energy spectrum and timing properties observed during an outburst, as the source moves counter-clockwise tracing a “q” shape in the hardness intensity diagram~\citep[HID; e.g.][]{2005Ap&SS.300..107H, 2010LNP...794...53B, 2011BASI...39..409B}, the outburst is generally categorized into four spectral–timing states: Low Hard State (LHS), Hard Intermediate State (HIMS), Soft Intermediate State (SIMS) and High Soft State (HSS). At the onset of an outburst, the source enters the LHS. During this state, the energy spectrum is dominated by the PL component. Strong broadband noise and Type-C QPOs are typically observed, with a high total fractional rms amplitude of $\sim 30$ percent in the $\sim 2-15$~keV band~\citep{1997ApJ...479..926M, 2005Ap&SS.300..107H, 2011MNRAS.410..679M}. Normally, as the outburst progresses, the source transitions into the intermediate states near the peak of the outburst light curve. In these states—which include the HIMS and the SIMS—the accretion disc emission becomes significant, accompanied by an increase in the temperature of the disc blackbody component. The timing properties in the HIMS are a continuation of those in the LHS, with an increase of the characteristic frequencies of the variability components and a decrease of the total fractional rms amplitude~\citep{2001ApJS..132..377H, 2005Ap&SS.300..107H, 2011MNRAS.410..679M}.  When the source transitions from the HIMS to the SIMS, Type-C QPOs are replaced by Type-B QPOs and the total fractional rms amplitude drops to below $\sim $10 percent~\citep{2005Ap&SS.300..107H, 2011MNRAS.410..679M, 2015MNRAS.447.2059M}.
When the source enters the HSS, the energy spectrum becomes dominated by the thermal emission from the accretion disc, while the PL component becomes weak or even undetectable~\citep{1997ApJ...479..926M, 2005Ap&SS.300..107H, 2010LNP...794...53B, 2024MNRAS.530..929J}. Sometimes, Type-A QPOs are detected during this state~\citep{2023MNRAS.526.3944Z}.

\subsection{Type-B QPO and jet ejections}

In the $\sim2-15$~keV band, the rms amplitude of Type-C QPOs varies from a few percent up to 20 percent~\citep{2002ApJ...564..962R, 2005ApJ...629..403C, 2015MNRAS.447.2059M}. These QPOs are usually narrow with quality factors $Q \geq 8$ ($Q=\nu/$FWHM, where $\nu$ is the centroid frequency and FWHM is the full width at half maximum of the QPO), exhibit variable frequencies ranging from 0.1 to 30 Hz, and are superimposed on a strong broadband noise~\citep[e.g.][]{2016AN....337..398M}.
For Type-B QPOs, the fractional rms amplitude is typically up to 5 percent in the $\sim 2-15$~keV band~\citep[e.g.][]{2019NewAR..8501524I}, with $Q \geq 6$ and variable frequencies in the range of $\sim 4-6$~Hz, accompanied by weak broadband noise variability.

The fractional rms amplitudes of both Type-C and Type-B QPOs initially increase with energy, flattening at energies above $10-20$~keV~\citep[e.g.;][]{2018ApJ...866..122H, 2020MNRAS.494.1375Z, 2022ApJ...938..108L}.
The energy-dependent phase-lag spectra of Type-C QPOs show significant evolution as the QPO frequency increases, transitioning from hard to soft lags~\citep[e.g.;][]{2020MNRAS.494.1375Z, 2025A&A...697A.229R}. A hard (soft) lag indicates that high- (low-) energy photons lag behind low- (high-) energy photons.
Type-B QPOs, on the other hand, usually show ``U''-shaped phase-lag spectra~\citep[e.g.;][]{2020MNRAS.496.4366B, 2021MNRAS.501.3173G, 2023MNRAS.525..854M, 2023MNRAS.519.1336P, 2023MNRAS.520.5144Z}.

Rapid and sudden transitions between Type-C and Type-B QPOs have been observed in BHXBs~\citep{2004A&A...426..587C, 2020ApJ...891L..29H, 2022ApJ...938..108L}.
\citet{2020ApJ...891L..29H} reported that in MAXI~J1820$+$070 a Type-C QPO, with a centroid frequency of $\sim 8$~Hz, abruptly transitions to a Type-B QPO with a centroid frequency of 4.5 Hz, accompanied by a sudden decrease in the strength of the broadband noise. A similar sudden transition was observed in MAXI~J1348$-$630~\citep[Figure 5 of][]{2022ApJ...938..108L}. In these cases, the power spectra exhibit a sudden transition from one pattern to another. 
Type-B and Type-C QPOs have previously been detected simultaneously in only two cases: during oscillations in the X-ray light curve of the peculiar source GRS 1915$+$105~\citep{2008MNRAS.383.1089S} and in GRO J1655$-$40, when it was in the atypical ultra-luminous state~\citep{2012MNRAS.427..595M, 2023MNRAS.525..221R}. So far, no instances of simultaneous Type-B and Type-C QPOs have been reported in any of the canonical accretion states, LHS, HIMS, SIMS, or HSS.

The appearance of Type-B QPOs, indicating the source enters the SIMS, is typically accompanied by a radio flare~\citep{2009MNRAS.396.1370F, 2012MNRAS.421..468M, 2019ApJ...883..198R, 2020ApJ...891L..29H}.
However, studies by \citet{2022MNRAS.511.4826C, 2024MNRAS.533.4188C} suggest that the correlation between Type-B QPOs and jet ejections is not always clear.
This radio flare is the result of discrete jet ejection events~\citep{2005ApJ...632..504C, 2012MNRAS.421..468M, 2019ApJ...883..198R, 2025ApJ...984L..53W}. A good example of the appearance of type-B QPOs at the same times of discrete jet ejections is reported by~\citep{2020ApJ...891L..29H} in MAXI~J1820$+$070.
Thanks to the potential of the power- and cross-spectra joint-fit method~\citep{2024MNRAS.527.9405M} to reveal weak variability components, in this paper we could reconsider the connection between the appearance of the Type-B QPO and the transition into the SIMS.

Several models have been proposed to reproduce the Type-C QPO and its properties, including general-relativistic Lense-Thirring precession~\citep{2009MNRAS.397L.101I, 2015MNRAS.446.3516I, 2016MNRAS.461.1967I, 2017MNRAS.464.2979I}, a resonance between the
disc and the corona~\citep[][]{2022A&A...662A.118M, 2022MNRAS.515.2099B}~\citep[see also;][]{2020MNRAS.492.1399K, 2021MNRAS.501.3173G}, and instabilities between a jet-emitting and standard accretion disc~\citep{2022A&A...660A..66F}. Regarding the origin of the Type‑B QPO, \citet{2016MNRAS.460.2796S} performed a phase-resolved spectroscopic study of the Type-B QPO in GX 339$-$4 and suggested that this QPO can be explained by a precessing jet~\citep[see also;][]{2020A&A...640L..16K}.

\subsection{Swift J1727.8$-$1613}

Swift J1727.8$-$1613 was first detected by Swift/BAT when the source went into a bright X-ray outburst in August 2023~\citep{2023GCN.34537....1P}. This source was later dynamically confirmed as a BHXB~\citep{2025A&A...693A.129M}.  During the outburst the X-ray photon count rate of the source rose rapidly and then decayed slowly~\citep{2024MNRAS.531.4893M, 2024MNRAS.529.4624Y}, followed by a series of X-ray flares~\citep{2024MNRAS.529.4624Y}. Strong Type-C QPOs were observed in the LHS and HIMS of the outburst in Swift J1727.8$-$1613~\citep[][]{2024MNRAS.531.4893M, 2024MNRAS.529.4624Y, 2025A&A...699A...9J}. The Type-C QPO frequency increased and then stabilized at $\sim 1.0-1.5$~Hz as the source entered the initial rise and plateau phases~\citep[][]{2024MNRAS.531.4893M, 2024MNRAS.529.4624Y, 2025A&A...699A...9J}. During the so-called flare state, The Type-C QPO frequency varied in a frequency range of $2-10$~Hz~\citep[][]{2024MNRAS.529.4624Y, 2025A&A...699A...9J}, accompanied by multiple radio flares~\citep{2024ApJ...968...76I, 2025ApJ...984L..53W}.

In this paper, we focus on the bright soft X-ray flare on 2023 September 19 (MJD 60206) reported by~\citet{2025ApJ...984L..53W}, and study the evolution of the variability during this flare using the method proposed by~\citet{2024MNRAS.527.9405M}. The paper is organized as follows: In Section~\ref{sec:OBSERVATION AND DATA REDUCTION} we describe the data reduction and analysis, in Section~\ref{sec:results} we show the results from the timing analysis, and in Section~\ref{sec:discussion} we discuss our results.

\section{Observations and data reduction}
\label{sec:OBSERVATION AND DATA REDUCTION}

We use the \textit{Insight}-HXMT data analysis software \textit{Insight}-HXMTDAS (version 2.06) and the latest CALDB files (version 2.07) to process the data, using only data from the small-FOV detectors. Here we analyze the \textit{Insight}-HXMT data of Swift J1727.8$-$1613 taken in the period September 17-22, 2023.
We create good time intervals (GTI) with the following selection criteria: pointing offset angle $< 0.04^{\circ}$, elevation angle $> 10^{\circ}$, geomagnetic cutoff rigidity $> 8$~GV, and the satellite being at least $300$~s away from the crossing of the South Atlantic Anomaly (SAA). 
We estimate the background with the {\sc lebkgmap, mebkgmap}, and {\sc hebkgmap} tasks in \textit{Insight}-HXMTDAS.

\subsection{Timing analysis}
\label{sec:Timing analysis}

We use GHATS\footnote{http://astrosat-ssc.iucaa.in/uploads/ghats\_home.html} to compute the Fast Fourier Transform to produce power-density spectra (PDS) in different energy bands and cross spectra (CS) of pairs of energy bands, for each individual GTI.
We compute the LE $2.0-10.0$~keV PDS, the HE $28-200$~keV PDS and the CS of photons in the HE $28-200$~keV band with respect to those in the LE $2.0-10.0$~keV band for each segment of 32~s, with a Nyquist frequency of 500~Hz. We then average those to get a PDS and CS per GTI\footnote{We use only the GTIs during which both the LE and HE instruments collected data simultaneously. and discard those GTI that are shorter than 320 seconds.}.
We rebin the averaged PDS and real and imaginary parts of the CS in frequency by a factor $\approx 1.012 = 10^{1/200}$ to increase the signal-to-noise ratio further but maintaining a good frequency resolution.
The PDS and the real and imaginary parts of the CS are normalized to fractional rms-squared units~\citep{1990A&A...230..103B}. The Poisson noise is estimated from the average power above 100 Hz, where no source variability is observed, and subtracted. The background count rate is considered when we compute the rms normalization.

We use Xspec~v.12.14.0~\citep{1996ASPC..101...17A} to fit the PDS and the real and imaginary parts of the CS.
We fit the LE $2-10$~keV PDS and HE $28-200$~keV PDS, as well as the real and imaginary parts of the corresponding CS simultaneously, using a combination of Lorentzian functions, as proposed by~\cite{2024MNRAS.527.9405M}. When we finished the fits, we calculate the 1$\sigma$ errors of the parameters in Xspec. We consider a Lorentzian component to be significant only if it exceeds a 3$\sigma$ significance level in at least one of the LE PDS, the  HE PDS or the CS. In Figs.~\ref{fig:flare_par},~\ref{fig:flare_rms},~\ref{fig:flare_hfd} and Table~\ref{tab:rep_obs} we show the parameters of the fit with 1$\sigma$ errors.

To generate fractional-rms and phase lag spectra of the QPOs, we divide the full energy range of the three \textit{Insight}-HXMT instruments into 12 bands. The separate energy bands are LE 1.0$-$1.5 keV, LE 1.5$-$2.0 keV, LE 2.0$-$2.6 keV, LE 2.6$-$4.8 keV, LE 4.8$-$7 keV, LE 7$-$11 keV, ME 7$-$11 keV, ME 11$-$15 keV, ME 15$-$35 keV, HE 25$-$35 keV, HE 35$-$48 keV, HE 48$-$100 keV. We select a time resolution of 3~ms (corresponding to a Nyquist frequency of $\sim 167$ Hz.) and a segment length of $\sim$ 25~s to produce the PDS and the CS. We use the LE 2.6$-$4.8 keV band as reference and each of the other bands as subject to produce the corresponding CS. In this manner, we obtain 12 PDS and 11 CS.
We fit jointly the 12 PDS and the real and imaginary parts of the 11 CS using a constant phase-lag model~\citep{2024MNRAS.527.9405M}. We also link the centroid frequency and FWHM for each Lorentzian function across the different spectra in the fit.
We subsequently calculate the 3-$\sigma$ confidence range of the parameters using the Markov chain Monte-Carlo algorithm (MCMC). 
The Goodman-Weare algorithm is applied for a total of 200000 samples and a burn-in phase of 100000 to 1000000 to let the chain reach a steady state.

\section{Results}
\label{sec:results}

\subsection{X-ray flare}
\label{sec:Flare in X-ray and radio}

\begin{figure}
    \vspace{-5mm}
	\includegraphics[width=\columnwidth]{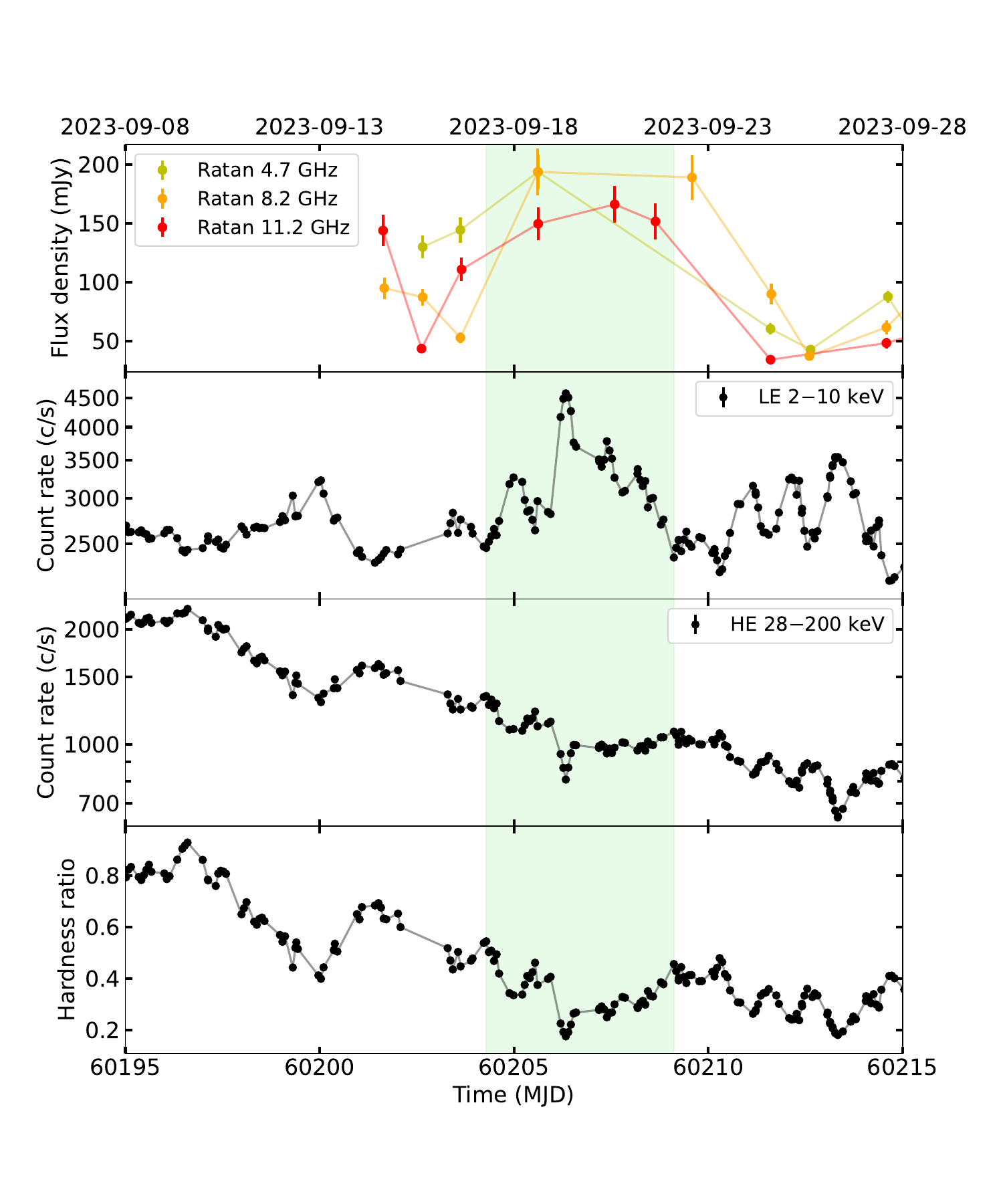}
	\vspace{-12mm}
    \caption[]{Ratan radio and \textit{Insight}-HXMT X-ray observations of Swift J1727.8$-$1613 in the period MJD 60195-60215. Each point corresponds to an individual GTI. From top to bottom, the panels show, respectively, the Ratan radio fluxes at three frequencies~\citep{2024ApJ...968...76I}, the LE $2-10$ keV light curve, the HE $28-200$ keV light curve and the corresponding hardness ratio.  The green shaded region indicates the observations we use in this paper.} 
    \label{fig:flare_light}
\end{figure}

\begin{figure}
    \vspace{-8mm}
	\includegraphics[width=\columnwidth]{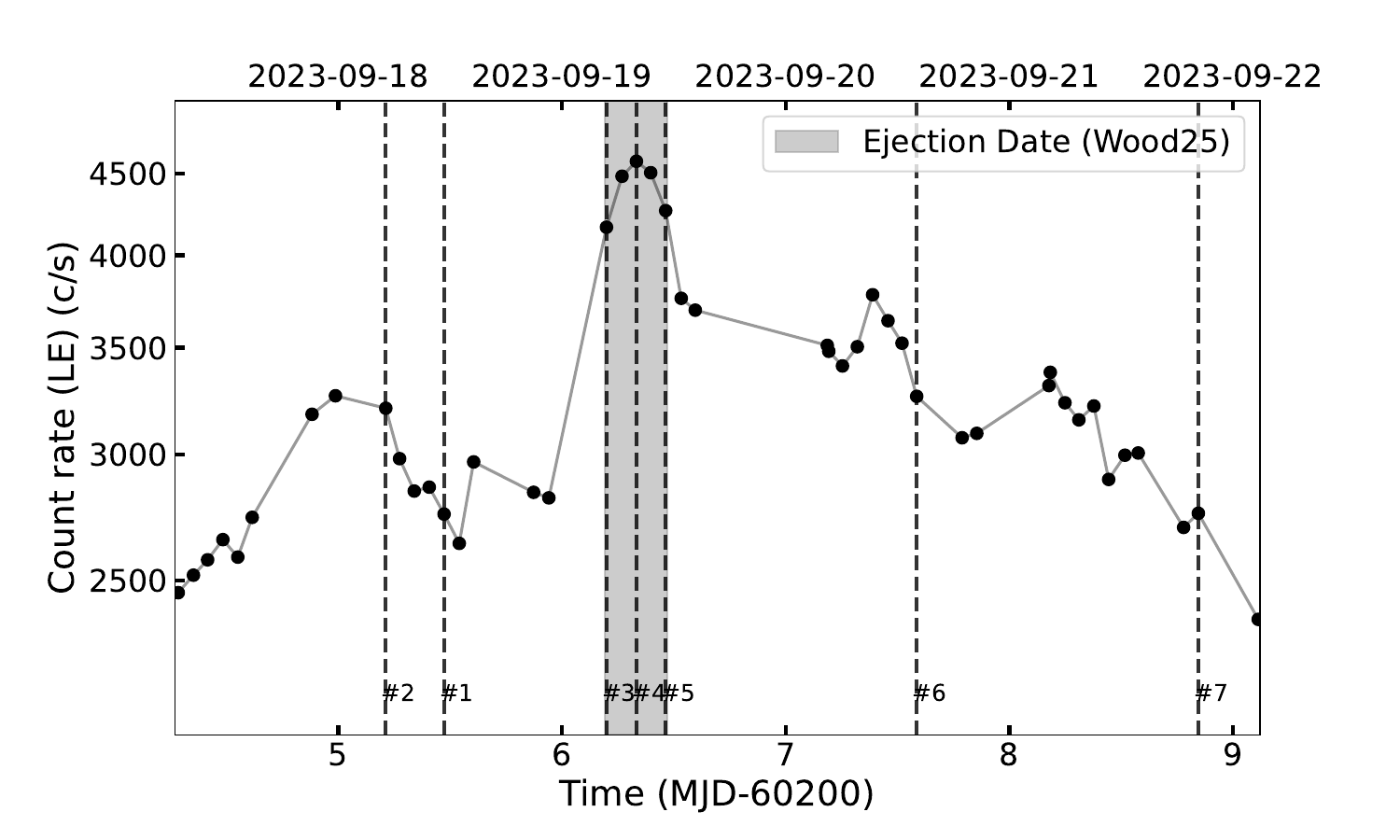}
	\vspace{-5mm}
    \caption{Zoom-in of the LE light curve of Swift J1727.8$-$1613 between MJD 60204 and 60209 in Fig.~\ref{fig:flare_light}. The gray shaded area marks the time, considering the uncertainty, of the jet ejections reported by~\citet{2025ApJ...984L..53W}, where a simultaneous soft X-ray flare occurs. The dashed vertical lines indicate the seven observations shown in Table~\ref{tab:rep_obs} and Fig.~\ref{fig:flare_representative_observations}.}
    \label{fig:flare_light_zoomin}
\end{figure}

In Fig.~\ref{fig:flare_light} we present the Ratan radio flux~\citep{2024ApJ...968...76I}, the LE 2.0$-$10.0~keV and HE 28.0$-$200~keV background-subtracted light curves, as well as the hardness ratio of Swift J1727.8$-$1613, spanning from MJD 60195 to MJD 60215. The hardness ratio is defined as the ratio between the background-subtracted count rate in the HE 28$-$200~keV and the LE 2$-$10~keV bands. The Ratan radio flux measurements are provided at three frequencies: 4.7 GHz (red), 8.2 GHz (orange), and 11.2 GHz (yellow).
The green shaded region marks the observations we used in this paper, between MJD 60204 to MJD 60209, lasting $\sim 5$ days.
In this period, the LE 2.0$-$10.0~keV photon count rate increases from $\sim 2500 $~cts/s to $\sim 4500 $~cts/s. In contrast, the HE 28$-$200~keV count rate decreases from $\sim 1300 $~cts/s to $\sim 800 $~cts/s. As a result, the hardness ratio decreases from 0.54 to 0.18 during the observations.
Simultaneously, the Ratan radio fluxes increase from $\sim 50$ mJy to $\sim 200$ mJy, although there are 2–6 day gaps between adjacent Ratan observations.

In Fig.~\ref{fig:flare_light_zoomin} we present a zoom-in version of the LE light curve of the observations we use. In addition, we mark the ejection date (MJD 60206.19$-$60206.47) of jet knots reported by~\citet{2025ApJ...984L..53W} with the gray region. Just prior to the ejection date, the LE $2-10$ keV light curve increases rapidly from $\sim 2800$ cts/s to $\sim4200$ cts/s, forming a soft flare in the LE light curve. As a result, the flare and jet ejections occur simultaneously on 2023 September 19 (MJD 60206), as reported by~\citet{2025ApJ...984L..53W}. The flare lasts $\sim 7$ hours.

\begin{table*}
	\centering
	\caption{Seven representative \textit{Insight}-HXMT observations of Swift J1727.8$-$1613, marked with black vertical dashed lines in Fig.~\ref{fig:flare_light_zoomin}. The position is relative to the flare on MJD 60206.}
	\label{tab:rep_obs}
    \scalebox{0.90}{
	\begin{tabular}{llllllllllll}
        \hline
        &\multirow{3}{*}{Obs ID} & \multirow{3}{*}{HR} &  & & \multirow{3}{*}{Position} &  \multicolumn{3}{c}{Type-C QPO} &  \multicolumn{3}{c}{Type-B QPO}\\
        Sequence& &  &  Obs time & Number of &   & Frequency & Q & rms (\%) & Frequency  & Q & rms (\%)\\ 
        number$^{*}$& &  & (MJD) &  segments &   &  (Hz) &  & in HE & (Hz)&  & in HE\\ 
        \hline
        \#1&P061433801903\_02 & 0.42 & 60205.47 & 17 & Before& $3.39\pm0.01$ & 12.6& $14.2 \pm 0.6$& $3.9 \pm 0.1$& 2.3& $9 \pm 1$\\
        \#2&P061433801901\_03 & 0.34 & 60205.21 & 15  & Before& $4.26\pm0.01$ &  19.0& $12 \pm 1$&$4.61 \pm 0.05$&  4.7& $11 \pm 1$ \\
        \#3&P061433802002\_03 & 0.23 & 60206.20 & 30 & Beginning& $6.67 \pm 0.03$& 6.7 & $9 \pm 1$&$7.41_{-0.04}^{+0.06}$& 6.7& $12.6^{+1.0}_{-0.4}$\\
        \#4&P061433802003\_02 & 0.18 & 60206.33 & 18 & Peak & $-$ & $-$ &$\leq1.5^{\dagger}$&$9.12\pm0.02$& 13.0& $14.4 \pm 0.4$\\
        \#5&P061433802004\_01 & 0.22 & 60206.46 & 28 & End& $6.84 \pm 0.03$& 7.6& $9 \pm 1$&$7.50 \pm 0.06$& 7.0&$13 \pm 1$\\
        \#6&P061433802104\_02 & 0.30 & 60207.59 & 27 & After& $4.92 \pm 0.01$& 12.3& $12 \pm 1$&$5.31 \pm 0.06$& 5.6& $10.6 \pm 0.8$\\
        \#7&P061433802206\_04 & 0.38 & 60208.84 & 20 & After& $3.97\pm0.01$& 12.4& $12.8 \pm 0.6$& $4.29_{-0.07}^{+0.08}$& 2.4& $9 \pm 1$\\
        \hline
        \multicolumn{12}{l}{$^{*}$ The observations are ordered according to the 2–10 keV intensity (rising – peak – declining), or equivalently, according to the} \\ 
        \multicolumn{12}{l}{hardness ratio (decreasing – minimum – increasing).} \\
        \multicolumn{12}{l}{ $^{\dagger}$ Type-C QPO is not detected and we give 95\% confidence upper limit of the fractional rms, $\leq1.8\%$ in the LE $2-10$ keV band}\\
         \multicolumn{12}{l}{and $\leq1.5\%$ in the HE $28-200$ keV band.}\\
	\end{tabular}
    }
\end{table*}

Seven representative \textit{Insight}-HXMT observations are marked with black vertical dashed lines in Fig.~\ref{fig:flare_light_zoomin}. Their observation times, hardness ratios and number of segments of 32 seconds used to generate the PDS are recorded in Table~\ref{tab:rep_obs}. The first two observations are before the flare, the next three are during the flare and the last two are after the flare.

\begin{figure}
    \vspace{-5mm}
	\includegraphics[width=\columnwidth]{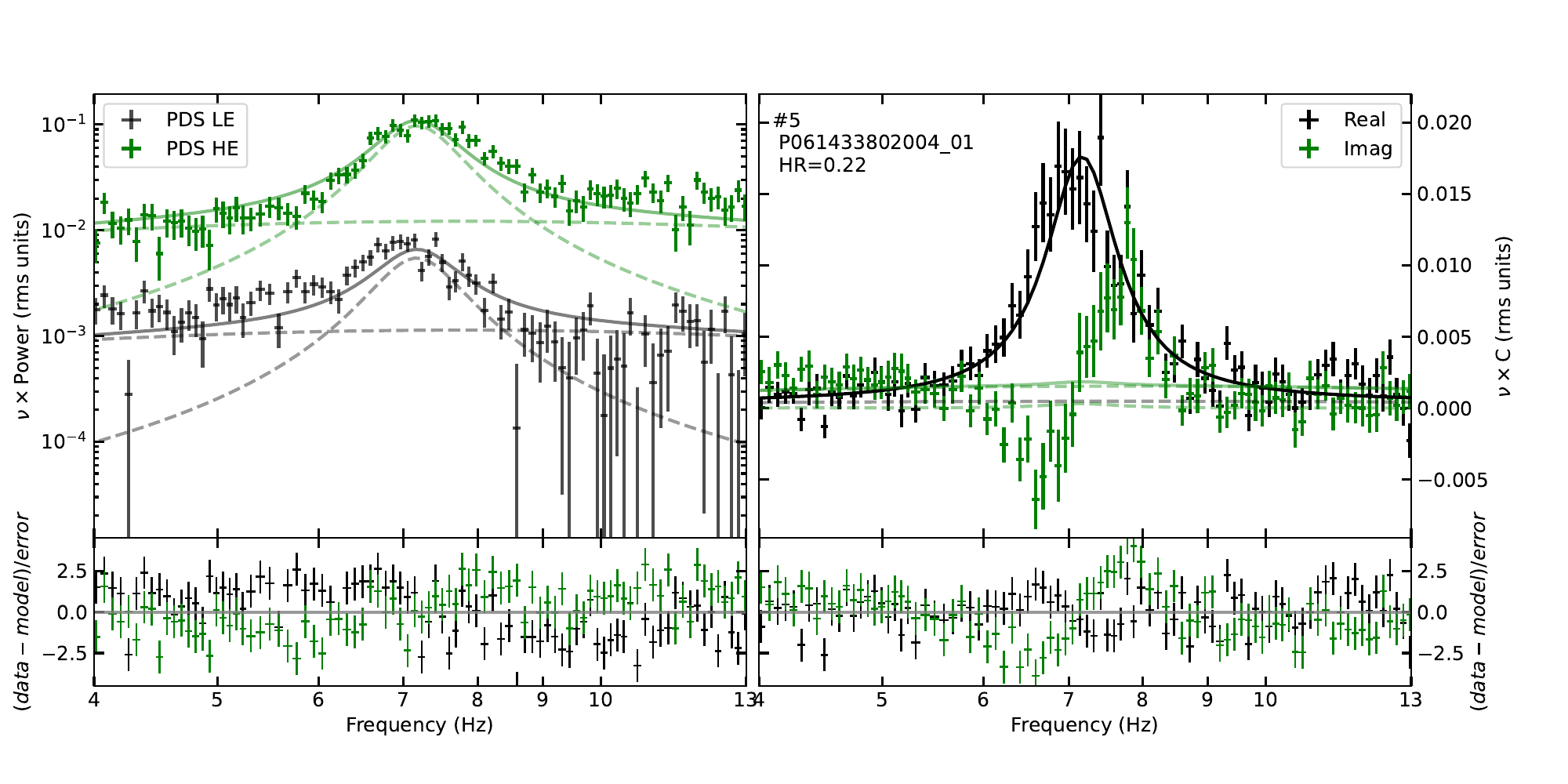}
	\vspace{-8mm}
    \caption{PDS and the CS of Obs \#5 in Table~\ref{tab:rep_obs}. Left panels: LE $2-10$ keV PDS (black) and HE $28-200$ keV PDS (green). Right panels: The real (black) and imaginary (green) parts of the CS between the HE and LE data.} 
    \label{fig:p2002_imag}
\end{figure}

In Fig.~\ref{fig:p2002_imag} we present the LE $2-10$ keV and HE $28-200$ keV PDS and the real and imaginary parts of the CS between the HE and LE data of Obs \#5 in Fig~\ref{fig:flare_light_zoomin} and Table~\ref{tab:rep_obs}. 
The LE and HE PDS peak at different frequencies: the peak in the HE PDS occurs at a slightly higher frequency than that in the LE PDS. 
The real part of the CS also shows an asymmetric QPO profile with a right-hand wing. 
This QPO has been reported by~\citet{2025ApJ...984L..53W}, and the asymmetry of the QPO profile is apparent in Figure 8 of ~\citet{2025MNRAS.tmp..712B}, although they did not mention it.
Interestingly, the imaginary part shows a very narrow negative peak, immediately followed by a very narrow positive peak at a slightly higher frequency, suggesting the presence of an additional variability component at that frequency, in addition to the Type-C QPO. This flip pattern of the imaginary part appears in most of the observations of Fig~\ref{fig:flare_light_zoomin}, except for those taken at the peak of the flare, where the imaginary part shows only a positive peak at the QPO frequency (see Fig.~\ref{fig:flare_representative_observations} and Section~\ref{sec:Dynamical cross spectrum}). 
We fit the LE and HE PDS, as well as the real and imaginary parts of the CS with 2 Lorentzians in the range 2$-$20 Hz; one Lorentzian fits the broadband signal and the other fits the QPO, with the frequencies and widths of the Lorentzians linked across the four spectra and the normalizations free to vary in each spectrum. The fit with this simple model also shows that a single Lorentzian is not enough to fit the QPO peak.
We note that \citet{2025A&A...699A...9J} fitted this QPO feature with two separate Lorentzians (see their Fig. 6), but did not discuss this in detail.

A QPO feature with a centroid frequency that depends on energy was also reported in several sources~\citep[e.g.;][]{2010ApJ...710..836Q, 2013MNRAS.428.1704L, 2013MNRAS.433..412L, 2018MNRAS.474.1214Y}. In the case of an observation of GRS~1915$+$105~\citep{2010ApJ...710..836Q}, \citet{2024MNRAS.527.9405M} suggested that this phenomenon can be explained more economically, with fewer free parameters, if the QPO feature consists of two separate Lorentzians with the same centroid frequency and FWHM in all energy bands. We perform a similar experiment, and fit the $2-12$ Hz frequency range of the 12 energy-band PDS of Obs.~\#3, in which the apparent energy dependence of the QPO frequency is most pronounced. We find that the model with two separate Lorentzians having centroid frequencies and FWHM that are the same across all 12 energy bands provides a better fit and requires fewer parameters ($\chi^2/\mathrm{dof} = 1495.3/1362$) than the single-Lorentzian model with an energy-dependent frequency and FWHM ($\chi^2/\mathrm{dof} = 1504.8/1354$).

In the following sections, we investigate this additional variability component together with the Type-C QPO.

\subsection{Joint-fit of power and cross spectra}
\label{sec:Joint-fit of power and cross spectra}

We fit simultaneously the PDS and CS using the method proposed by~\citet{2024MNRAS.527.9405M}, assuming the constant phase-lag model described by those authors.
This method is based on the assumptions that: (i) the PDS can be described as a linear combination of Lorentzian functions (additive components); (ii) all the Lorentzians have the same centroid frequency and FWHM in each energy band; (iii) each Lorentzian is perfectly coherent in any two energy bands; (iv) any two Lorentzians are incoherent with one another.
Assumption (iv) only needs to apply over the time scales of the Lorentzians themselves, and is not affected if the frequencies or other properties of two or more of those Lorentzian are correlated over longer timescales.
In previous papers~\citep[e.g.;][]{2024MNRAS.527.9405M, 2025A&A...696A.128B, 2025A&A...696A.237F, 2025arXiv250507938B, 2025ApJ...990...43R, 2025A&A...699A...9J}, it was shown that for different sources and different instruments the model reproduces the PDS and the real and imaginary parts of the CS, and it predicts correctly the lags and the coherence function.

\begin{figure}
    \vspace{-20mm}
	\includegraphics[width=\columnwidth]{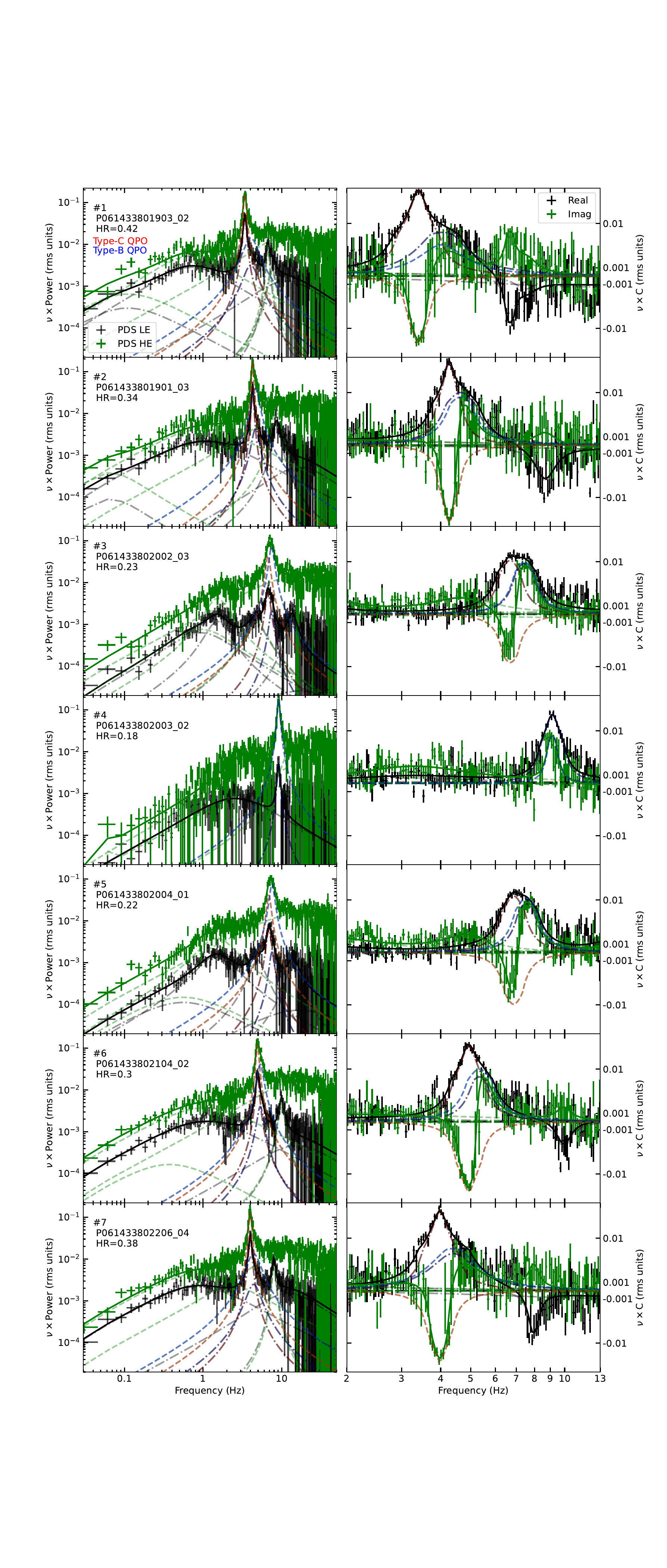}
	\vspace{-28mm}
    \caption{PDS and CS of the seven representative \textit{Insight}-HXMT observations marked with vertical lines in Fig.~\ref{fig:flare_light_zoomin}, and given in Table~\ref{tab:rep_obs}. Left panels: LE $2-10$ keV PDS (black) and HE $28-200$ keV PDS (green). Right panels: The real (black) and imaginary (green) parts of the CS. The individual Lorentzians are shown as dashed-dotted lines in the LE PDS and dashed lines in the HE PDS; in the CS, the dashed-dotted lines indicate the real part, and the dashed lines indicate the imaginary part. We highlight the Type-C (red) and Type-B (blue) QPOs in the PDS and the CS. Data with absolute values greater than or equal to $5\times10^{-3}$ are scaled logarithmically, while values with absolute values smaller than  $5\times10^{-3}$ are scaled linearly.} 
    \label{fig:flare_representative_observations}
\end{figure}

We present the LE $2-10$~keV (black) and the HE $28-200$~keV PDS (green) of the observations recorded in Table~\ref{tab:rep_obs} in the left panels of Fig.~\ref{fig:flare_representative_observations}. The right panels show the real (black) and  imaginary (green) parts of the CS of the light curve in the HE $28-200$~keV band with respect to that in the LE $2-10$~keV band. QPOs are clearly visible in both the PDS and CS.
We also show the contribution of each Lorentzian to the PDS and CS. We highlight the Type-C QPO, which consistently appears in the LHS and HIMS~\citep{2025A&A...697A.229R, 2025A&A...699A...9J}, in red in both the PDS and CS. 
We also highlight, in blue, the component revealed by the flip pattern of the imaginary part in Fig.~\ref{fig:p2002_imag}. This component initially appears as a shoulder of the Type-C QPO~\citep{1997A&A...322..857B, 2002ApJ...572..392B, 2020MNRAS.496.5262V, 2024MNRAS.527.9405M} in the PDS, which transitions from a broad to a narrow feature as the source approaches the soft flare on MJD 60206, ultimately becoming a canonical QPO as the source reaches the peak of the flare. After the peak of the flare, the width of the QPO increases again, as it once more becomes a shoulder of the Type-C QPO. This transition is clearly visible in the HE PDS. In observations P061433801903\_02, P061433801901\_03, P061433802104\_02 and P061433802206\_04, taken before and after the flare, this component looks like a shoulder of the Type-C QPO. In observations P061433802002\_03 and P061433802004\_01, corresponding to the beginning and end of the flare, respectively, this component becomes dominant in the HE PDS while the Type-C QPO dominates the LE PDS, as suggested by the misalignment of the peaks of the LE and HE PDS (see also Fig.~\ref{fig:p2002_imag}). In these two observations the amplitude of this component in the CS is larger than before and after the flare, leading to a more pronounced asymmetry in the real part and a sharper flip in the imaginary part. At the peak of the flare, in observation P061433802003\_02, the Type-C QPO disappears and this component is the only QPO present in the PDS and CS. 
Based on a number of properties of this component and the broadband PDS, which we  summarize in Section~\ref{sec:Identification of the Type-B QPO}, we identify this component as the Type-B QPO; therefore from now on we refer to this QPO as the Type-B QPO.
We record the parameters of the Type-C and Type-B QPOs in Table~\ref{tab:rep_obs}. Although we do not detect the Type-C QPO in observation P061433802003\_02, we estimate an upper limit of its fractional rms amplitude, which is  $\leq1.8\%$ in the LE $2-10$ keV band and $\leq1.5\%$ in the HE $28-200$ keV band (95\% confidence level), significantly lower than the rms amplitude of this QPO in other observations (see Table~\ref{tab:rep_obs}).

In Fig.~\ref{fig:flare_representative_observations_1} we display the rotated real and imaginary parts of the rotated CS~\footnote{As in \citet{2024MNRAS.527.9405M}, we rotate the cross vectors counterclockwise by $\pi/4$ rad, such that variability components with $\sim$ zero phase lags would have approximately equal real and imaginary parts. On the one hand, this stabilize the fits, and on the other hand it allows to plot the CS using a logarithmic scale, which improves the visibility of the component with a small imaginary part. Since this is equivalent to a rotation of the axes by $-\pi/4$ rad, this has no impact on the fit process \citep[see Appendix~\ref{sec:rotated cs} and][for details]{2024MNRAS.527.9405M}. }. 
The rotated CS suggests again that the Type-C and Type-B QPOs are independent variability components (see Appendix~\ref{sec:rotated cs}).

\begin{figure}
    \vspace{-5mm}
	\includegraphics[width=\columnwidth]{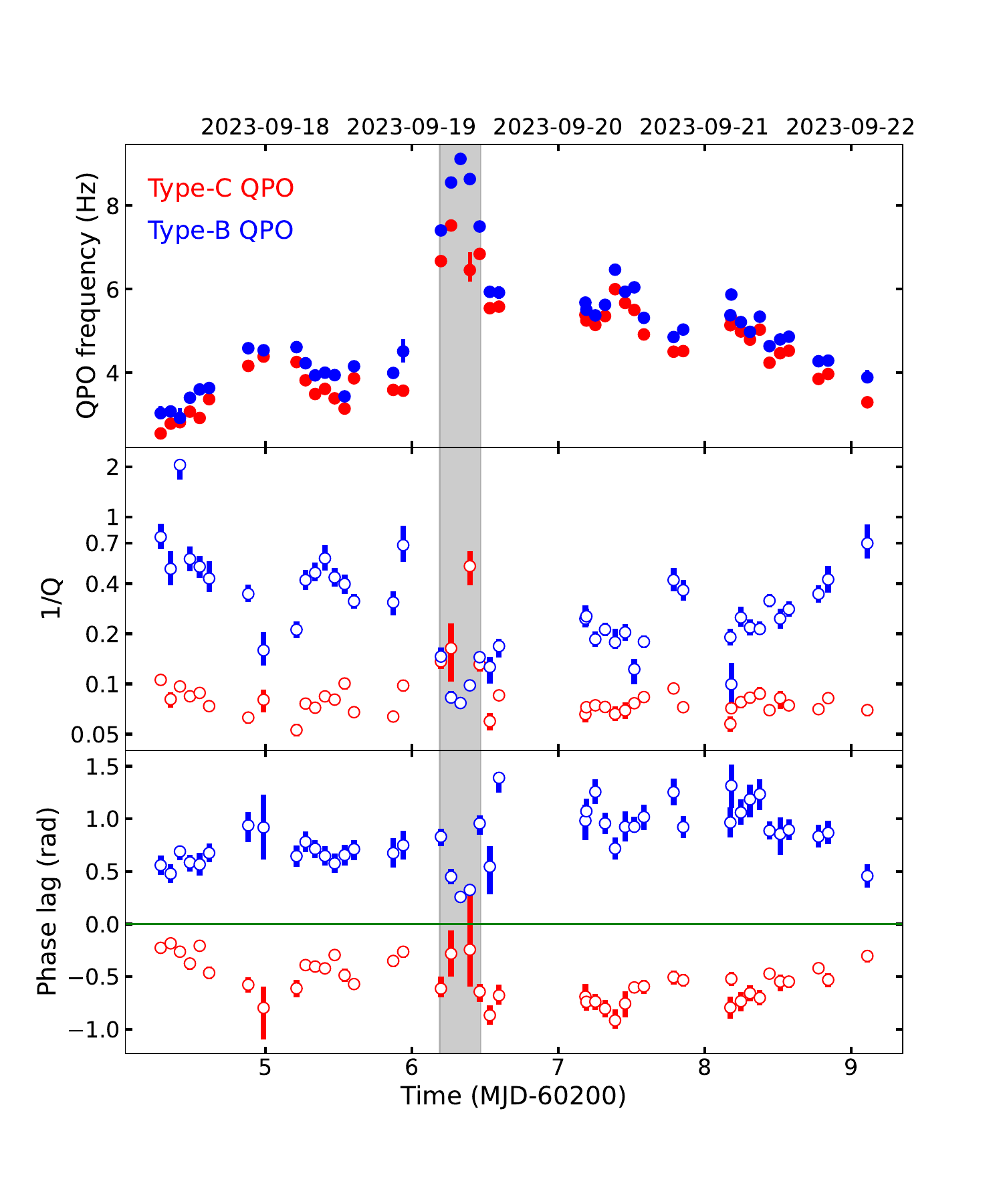}
	\vspace{-13mm}
    \caption{Parameters of the Type-C and Type-B QPOs of the observations of Swift J1727.8$-$1613 in Fig.~\ref{fig:flare_light_zoomin}. Top panel: the QPO frequency. Middle panel: $1/Q$. Bottom panel: the phase lag. The gray area marks the time of the soft X-ray flare and the jet ejections reported by~\citet{2025ApJ...984L..53W}.}
    \label{fig:flare_par}
\end{figure}

In Fig.~\ref{fig:flare_par} we present the evolution of the QPO frequencies, $1/Q$, and the phase lags of the Type-C QPO (red) and  Type-B QPO (blue), with $1\sigma$ errors. 
The frequency of the Type-B QPO is always slightly higher than that of the Type-C QPO. As the source approaches the peak of the flare, the frequencies of both QPOs increase, from $\sim$2.6 Hz to $\sim$7.5 Hz for the Type-C QPO, and from $\sim$3.1 Hz to $\sim$9.1 Hz for the Type-B QPO, the latter being comparable to the highest frequency of the Type-B QPO observed in GRO~J1655$-$40~\citep{2012MNRAS.427..595M}. The most rapid increase occurs at the beginning of the soft X-ray flare on MJD 60206.
Throughout the observations, $1/Q$ for the Type-C QPO remains approximately constant at $\sim $0.1, except for the observation at the peak of the flare where the Type-C QPO is not detected. In contrast, the $1/Q$ for the Type-B QPO decreases from 1 to 0.1. As a result, $1/Q$ of the Type-B QPO reaches the minimum values in the flare, which is $\sim 0.16-0.07$. This indicates that in the flare the quality factor of the Type-B QPO is $\sim6.4-13.7$, consistent with those ($\geq$6) of the Type-B QPOs observed in the SIMS of the other sources~\citep{2005ApJ...629..403C}.
Throughout the observations, the phase lags of the two QPOs differ significantly, by at least 3$\sigma$. The phase lag of the Type-C QPO remains consistently negative, varying between $-$1.0 rad and $\geq-0.1$ rad. In contrast, the phase lag of the Type-B QPO is positive, ranging from 1.5 rad to 0.2 rad.
We find that the phase lag of the Type-C QPO is anti-correlated with the QPO frequency, consistent with previous results~\citep[e.g.][]{2020MNRAS.494.1375Z}, while the lags of the Type-B QPO show a positive correlation with the QPO frequency~\citep[e.g.][]{2017MNRAS.466..564G}.
In Fig.~\ref{fig:flare_par_rot}, we compare the fit results with and without the rotation of the CS (see above) and find them to be consistent within 1$\sigma$ uncertainties. As expected~\citep{2024MNRAS.527.9405M}, the rotation does not affect the fit results.

\begin{figure}
    \vspace{-5mm}
	\includegraphics[width=\columnwidth]{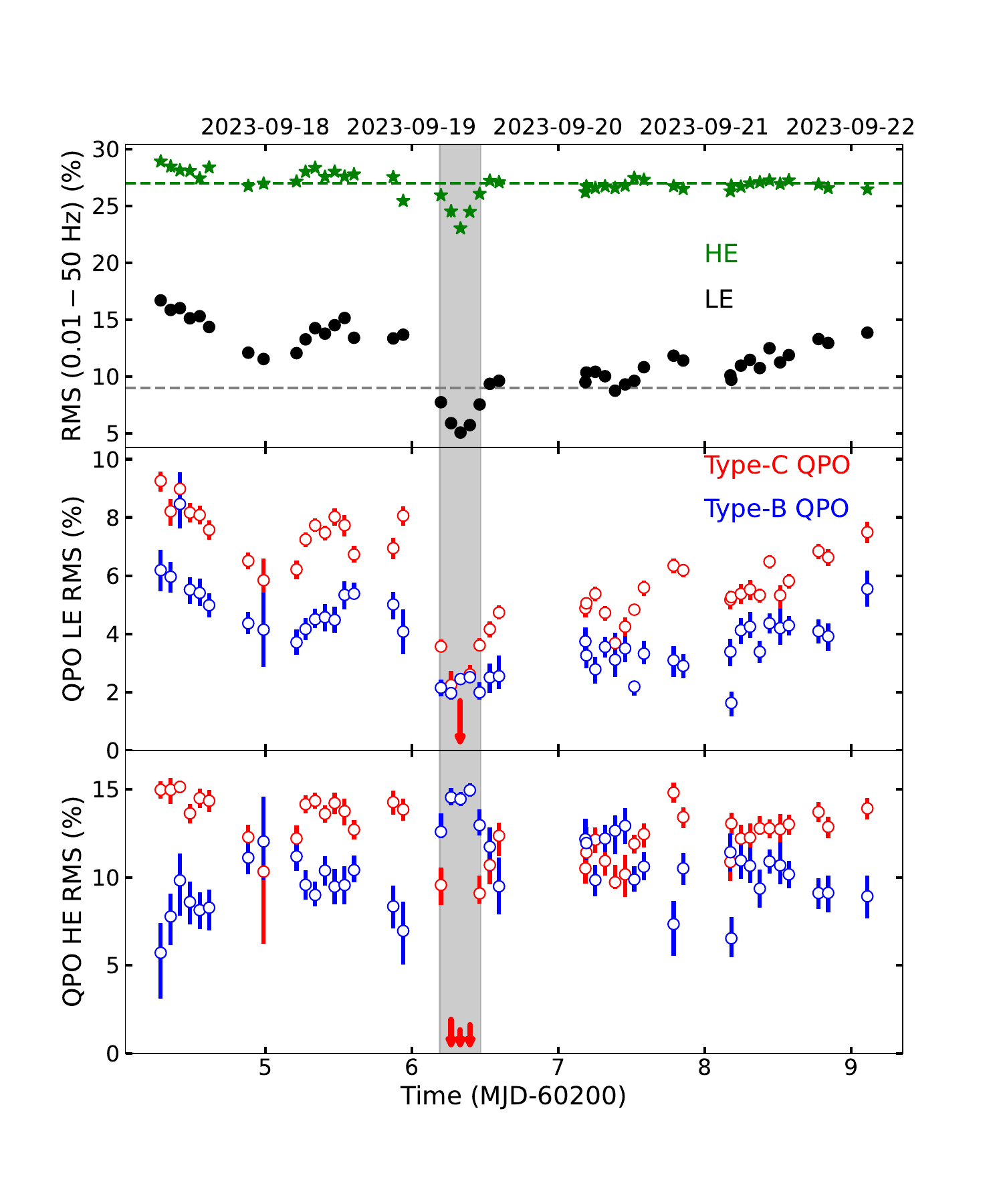}
	\vspace{-13mm}
    \caption{Top panel: Total ($0.01-50$~Hz) fractional rms amplitude of Swift J1727.8$-$1613 in the LE $2-10$ keV (black) and HE $28-200$ keV (green) bands. Middle panel: Fractional rms amplitudes of the Type-C (red) and Type-B (blue) QPOs in the LE $2-10$ keV band. Bottom panel: Fractional rms amplitudes of the Type-C and Type-B QPOs in the HE $28-200$ keV band. The gray area marks the time of the soft X-ray flare and the jet ejections reported by~\citet{2025ApJ...984L..53W}. The arrows indicate an upper limit.}  
    \label{fig:flare_rms}
\end{figure}

In the top panel of Fig.~\ref{fig:flare_rms} we present the total ($0.01-50$~Hz) fractional rms amplitudes for LE 2.0$-$10~keV (black) and HE $28-200$~keV (green), in the middle panel the fractional rms amplitudes of the Type-C (red) and Type-B (blue) QPOs in the LE 2.0$-$10~keV band, and in the bottom panel the fractional rms amplitudes of the Type-C (red) and Type-B (blue) QPOs in the HE $28-200$~keV band. As we mentioned before, and as we indicated in Table~\ref{tab:rep_obs}, the Type-C QPO is not detected in the observation at the peak of the flare on MJD 60206. Therefore, in the middle and bottom panels of Fig.~\ref{fig:flare_rms} we plot the upper limit of the fractional rms of the Type-C QPO.
In the flare, on MJD 60206, consistent with the time of the ejection of jet knots reported by~\citep{2025ApJ...984L..53W},  the broadband fractional rms amplitude drops to below 9\% in the LE band and below 27\% in the HE band. The total fractional rms in the LE 2.0$-$10~keV band during the drop is consistent with those observed in the SIMS of other sources~\citep[below 10\% in the $\sim 2-15$ keV;][]{2011MNRAS.410..679M}.
The fractional rms of the Type-C QPO in the LE band decreases from $\sim$ 10\% to $\sim$ 1\% as the source approaches the peak of the flare, reaching the minimum at the peak.
The fractional rms of the Type-B QPO in the LE band also decreases slightly, but not significantly from $\sim$ 6\% to $\sim$ 2\%, consistent with the 2$-$26 keV fractional rms of the Type-B QPOs observed in other sources~\citep{2015MNRAS.447.2059M}, as the source approaches the peak of the flare.
The fractional rms of the Type-C QPO in the HE band shows a significant drop during the flare, becoming not significant at the peak of  the flare.
In contrast, the fractional rms of the Type-B QPO in the HE band increases from $\sim$ 5\% to $\sim$ 15\% as the source approaches the peak of the flare. 
Consequently, the time (MJD 60206.20) at which the Type-B QPO becomes stronger than the Type-C QPO in the HE PDS aligns closely with the ejection time of jet knot 3 (MJD 60206.22$\pm$0.03) reported by \citet{2025ApJ...984L..53W}.

\begin{figure}
    \vspace{-8mm}
	\includegraphics[width=\columnwidth]{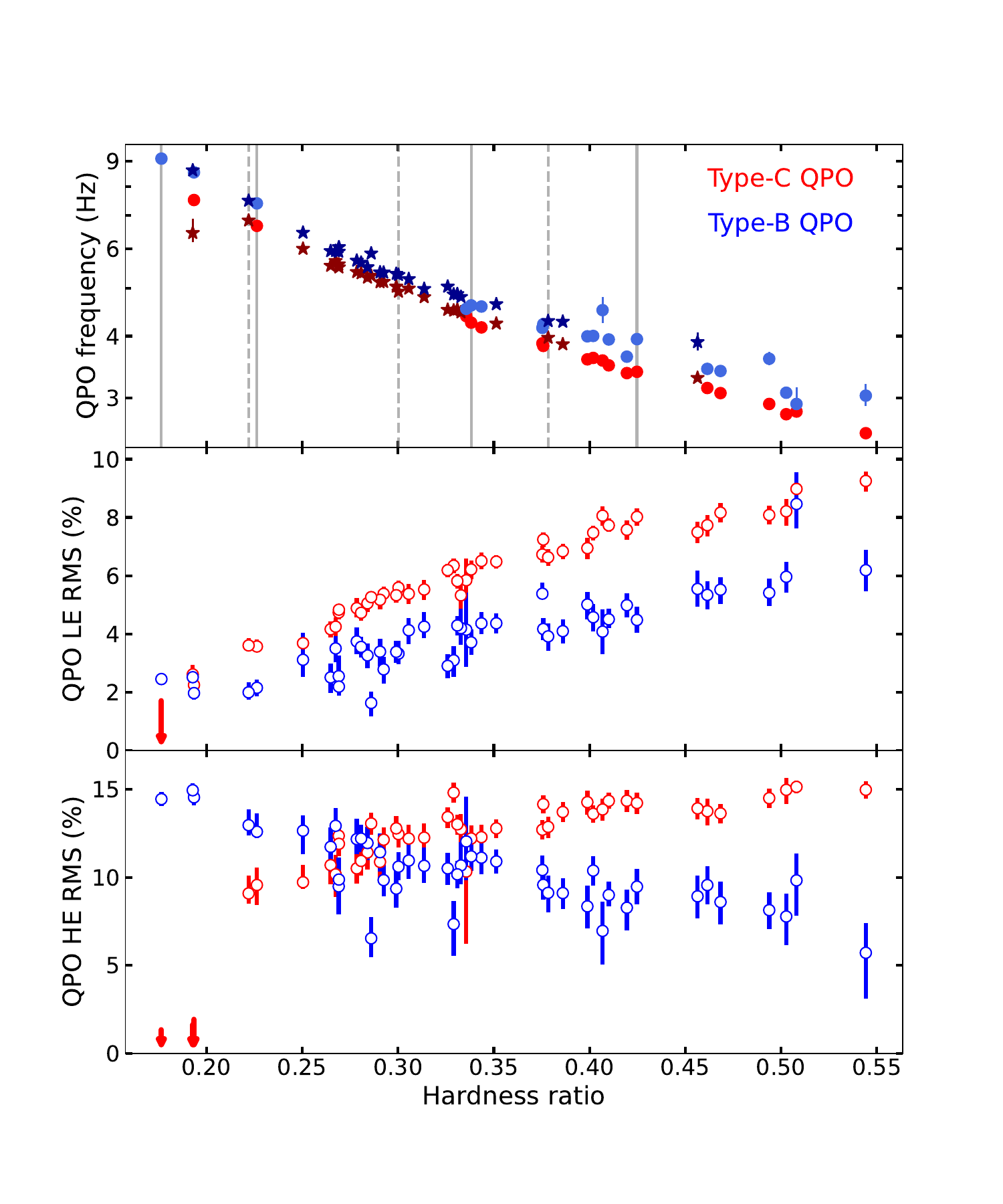}
	\vspace{-13mm}
    \caption{Frequency and fractional rms amplitude of the Type-C (red) and Type-B (blue) QPOs as a function of the hardness ratio. Top panel: The QPO frequency. Circles represent the rising phase of the LE light curve, while stars represent the decay phase. Vertical lines mark the representative \textit{Insight}-HXMT observations in Table~\ref{tab:rep_obs}. Notably, the Type-C QPO is not detected at the peak of the flare, which corresponds to the minimum hardness ratio. Middle panel: the fractional rms amplitude in the LE band. Bottom panel: the fractional rms amplitude in the HE band. The arrows indicate an upper limit.}  
    \label{fig:flare_hfd}
\end{figure}

In the first panel of  Fig.~\ref{fig:flare_hfd} we present the QPO frequencies as a function of the hardness ratio. The frequencies of both the Type-C and Type-B QPOs, $\nu_C$ and $\nu_B$, respectively, show an anti-correlation with the hardness ratio, indicating that both QPOs are correlated with each other on timescales of hours to days. 
We find that $\nu_B \sim 1.1\nu_C$, inconsistent with one QPO being a low-order harmonic of the other. We note that a 10:11 harmonic ratio, without a significant fundamental or low-order harmonic, has never been reported in this, or any other BHXB. 
On the other hand, this correlation is consistent with previous findings that the frequencies of all variability components, both QPOs and broad Lorentzians, are correlated with each other throughout an outburst~\citep[e.g.][]{1999ApJ...520..262P}.
While a similar correlation for the Type-C QPOs has been observed in BHXBs~\citep{2022NatAs...6..577M, 2022MNRAS.514.2891Z, 2022MNRAS.513.4196G, 2024MNRAS.527.7136B}, this is the first time that such a correlation has been reported for the Type-B QPOs. In the last two panels of Fig.~\ref{fig:flare_hfd} we present the fractional rms amplitudes of the Type-C and Type-B QPOs in the LE and HE bands, respectively. 
The fractional rms of the Type-C QPOs in both the LE and HE bands is positively correlated with the hardness ratio. For the Type-B QPOs the fractional rms in the LE band also shows a positive correlation with hardness. However, its fractional rms in the HE band is anti-correlated with hardness.

\subsection{Energy-dependent fractional-rms and phase-lag spectra}
\label{sec:Energy-dependent fractional-rms and phase-lag spectra}

\begin{figure}
    \vspace{-25mm}
	\includegraphics[width=\columnwidth]{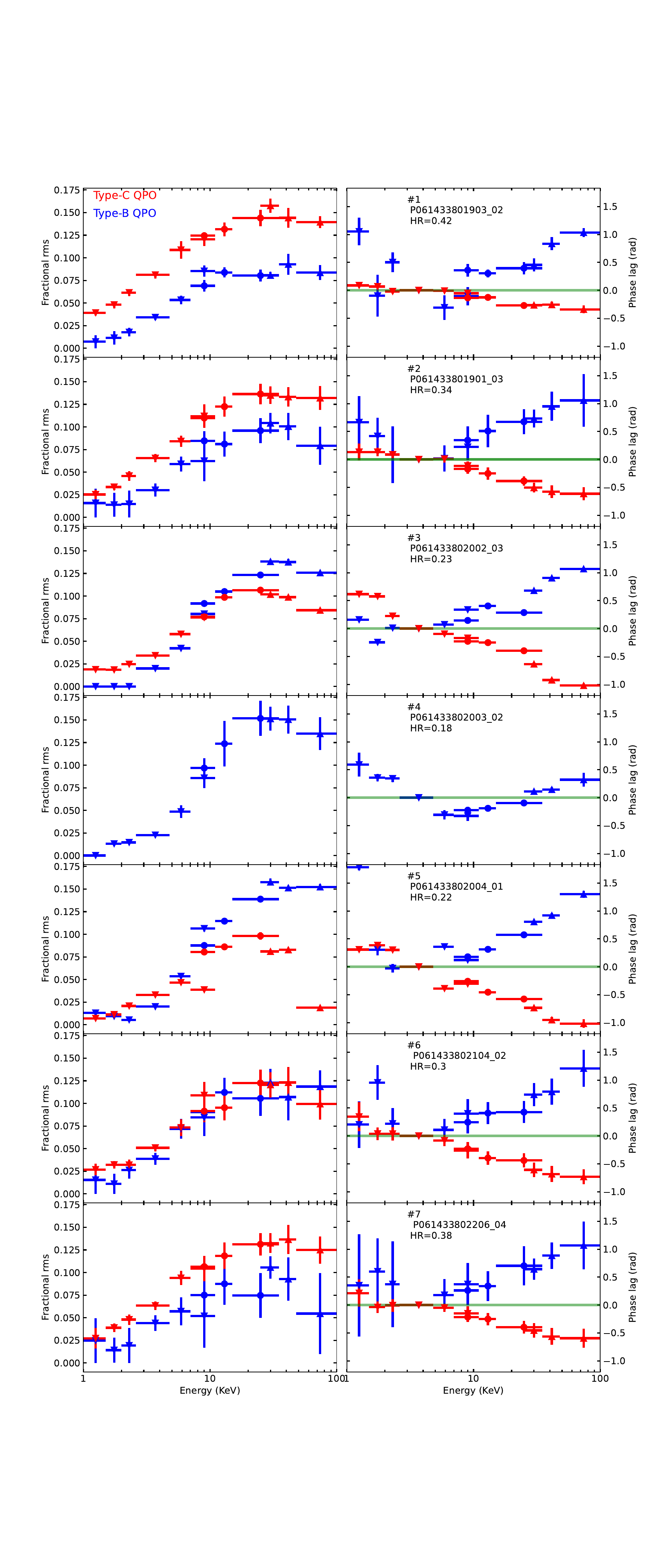}
	\vspace{-28mm}
    \caption{Fractional-rms (left panels) and phase-lag spectra (right panels) of the Type-C (red) and Type-B (blue) QPOs of Swift J1727.8$-$1613 for the representative \textit{Insight}-HXMT observations in Table~\ref{tab:rep_obs}.}  
    \label{fig:flare_representative_observations_spectra}
\end{figure}

In Fig.~\ref{fig:flare_representative_observations_spectra} we present the energy-dependent fractional-rms and phase-lag spectra of the Type-C and Type-B QPOs for the seven representative observations in Table~\ref{tab:rep_obs}. The fractional rms of Type-C QPO initially increases in the energy range of $1-15$~keV, and above 15~keV it either remains more or less constant for observations P061433801903\_02, P061433801901\_03, P061433802104\_02 and P061433802206\_04,  or decreases for observations P061433802002\_03 and P061433802004\_01. The maximum fractional rms of Type-C QPO decreases from $\sim$ 17\% to $\sim$ 10\% as the source approaches the peak of the LE light curve, where the Type-C QPO eventually disappears.
For the Type-B QPO the fractional rms also increases with energy in the $1-15$~keV band, and then remains more or less constant. In contrast to the Type-C QPO, the maximum fractional rms of the Type-B QPO increases from $\sim$ 8\% to $\sim$ 15\% as the source approaches the peak of the LE light curve, which is comparable to the case reported by~\citet{2022ApJ...938..108L}, where the fractional rms of the Type-B QPO reaches
 $\sim$ 11\% above $\sim$ 10 keV.
The phase lag of Type-C QPO decreases with energy in all observations. A soft lag is also reported by \citet{2025MNRAS.tmp..712B} when the frequency is between 2$-$6 Hz.
On the contrary, the phase lag of the Type-B QPO initially decreases with energy and then increases. Similar phase-lag trends for the Type-B QPO, a "U"-shaped phase lag spectrum, have also observed in other sources such as MAXI~J1348$-$630 \citep{2020MNRAS.496.4366B, 2021MNRAS.501.3173G}, MAXI~J1820$+$070 \citep{2023MNRAS.525..854M}, GX~339$-$4~\citep{2023MNRAS.519.1336P} and MAXI~J1535$-$571~\citep{2023MNRAS.520.5144Z}. In Swift J1727.8$-$1613 the minimum phase lag of the Type-B QPO occurs $\sim$ 3~keV, except for the observation P061433802003\_02 at the peak of the flare, where the minimum shifts slightly higher energies, $\sim 9$~keV.

\vspace{-4mm}
\subsection{Identification of the Type-B QPO}
\label{sec:Identification of the Type-B QPO}
\vspace{-1mm}

\begin{figure*}
    \vspace{-5mm}
	\includegraphics[width=\textwidth]{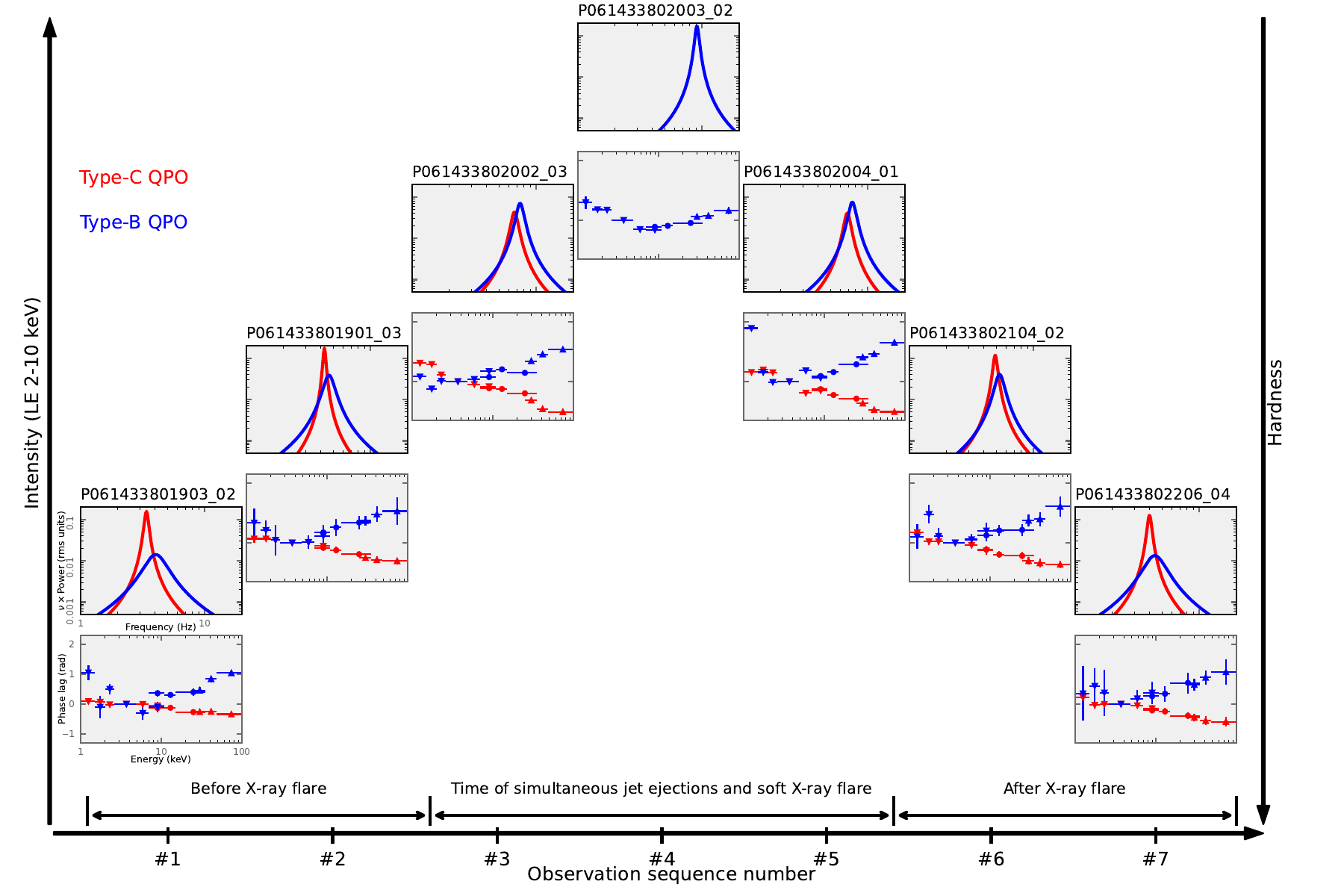}
	\vspace{-5mm}
    \caption{Evolution of the Type-C (red) and Type-B (blue) QPOs in the HE PDS and corresponding phase-lag spectra across seven observations in Table~\ref{tab:rep_obs} and Fig.~\ref{fig:flare_representative_observations}, ordered by observation sequence number along the horizontal axis. The x- and y-ranges are the same across all panels. The vertical axes represent increasing $2-10$ keV intensity (upward) and decreasing spectral hardness (downward). The profiles of the Type-C and Type-B QPOs are extracted from Fig.~\ref{fig:flare_representative_observations}, with all other data and Lorentzian components omitted for clarity. The phase lag spectra shown are identical to those presented in Fig.~\ref{fig:flare_representative_observations_spectra}. The first two observations occur before the soft X-ray flare on MJD 60206, the next three during the flare, and the final two after the flare. } 
    \label{fig:fig_flare_rep_qpo}
\end{figure*}

The imaginary part of the CS in Fig.~\ref{fig:p2002_imag} shows a variability component at a frequency slightly higher than that of the Type-C QPO, which is present in the data even before we apply the Lorentzian model described in Section~\ref{sec:Joint-fit of power and cross spectra}.
We show this component in blue in the PDS and CS in Fig.~\ref{fig:flare_representative_observations}. In Appendix~\ref{sec:coh} we present the coherence function and phase-lag spectra, indicating again that this component is an independent variability component, which appears as the shoulder of the Type-C QPO in the PDS of most of the observations.
In Fig.~\ref{fig:fig_flare_rep_qpo} we illustrate the smooth evolution of this component (blue) alongside the Type-C QPO (red) in the HE PDS. As we describe in Section~\ref{sec:Energy-dependent fractional-rms and phase-lag spectra}, the two features are clearly distinguished by their phase lag spectra: this variability component (blue) displays a characteristic “U”-shaped profile, whereas the Type-C QPO (red) shows a steadily decreasing trend with energy.
We can clearly see that, in the PDS, this component transitions from a relatively broad shoulder of the Type-C QPO to a regular QPO as the source approaches the peak of the soft X-ray flare on MJD 60206 (Fig.~\ref{fig:flare_representative_observations}; this evolution is depicted schematically in Fig.~\ref{fig:fig_flare_rep_qpo}).
At the peak of the flare, this QPO has replaced the Type-C QPO, and is the only narrow feature in the PDS (middle panel of Fig.~\ref{fig:flare_representative_observations}). After the peak of the flare, the Type-C QPO reappears, this other QPO becomes broader and it once more turns into a shoulder of the Type-C QPO in the PDS.
Furthermore, we also find that:
(i) the ``U''-shaped phase-lag spectrum of this component (Fig.~\ref{fig:flare_representative_observations_spectra}) is similar to those seen in other Type-B QPOs;
(ii) during the flare, the quality factor of this feature exceeds 6 (Fig.~\ref{fig:flare_par}), a typical value for the Type-B QPO~\citep{2005ApJ...629..403C}, while at the same time the total ($0.01-50$~Hz) fractional rms drops below 10\% in the $2-10$ keV band (Fig.~\ref{fig:flare_rms}), similar to the values observed in other sources in the SIMS when the Type-B QPO appears~\citep{2011MNRAS.410..679M};
(iii) during the flare, when this QPO dominates the PDS, the discrete radio jet ejections occur~\citep[Fig.~\ref{fig:flare_light_zoomin};][]{2025ApJ...984L..53W}, which is similar to what is observed in other sources during the transition from the HIMS to the SIMS when Type-B QPOs appear~\citep[e.g.;][]{2009MNRAS.396.1370F, 2020ApJ...891L..29H}.
These properties are consistent with those of Type-B QPOs reported in the literature, supporting the identification of this feature as the Type-B QPO in Swift J1727.8$-$1613. In addition, these properties indicate that the source enters the traditional SIMS during the flare on MJD 60206 (see a detailed discussion in Section~\ref{sec:Type-B QPOs and SIMS}).

In the case of Swift J1727.8$-$1613, the QPOs exhibit the following properties:
(i) the Type-B QPO is detected in all our observations during both the HIMS and SIMS, rather than being limited to the SIMS;
(ii) the Type-B QPO co-exists with the Type-C QPO during the HIMS;
(iii) the frequency of the Type-B QPO is anti-correlated with the hardness ratio;
(iv) the frequency of the Type-B QPO in Swift J1727.8$-$1613 is consistently higher than that of the Type-C QPO, in contrast to the only other case in which the Type-B and C were detected simultaneously~\citep[GRO J1655$-$40;][]{2012MNRAS.427..595M}. Since in other sources the Type-B and -C QPOs appear at different times, it is not possible to compare their frequencies to draw general conclusions.

\vspace{-6mm}
\subsection{Dynamical imaginary cross-spectrum}
\label{sec:Dynamical cross spectrum}

\begin{figure}
	\includegraphics[width=\columnwidth]{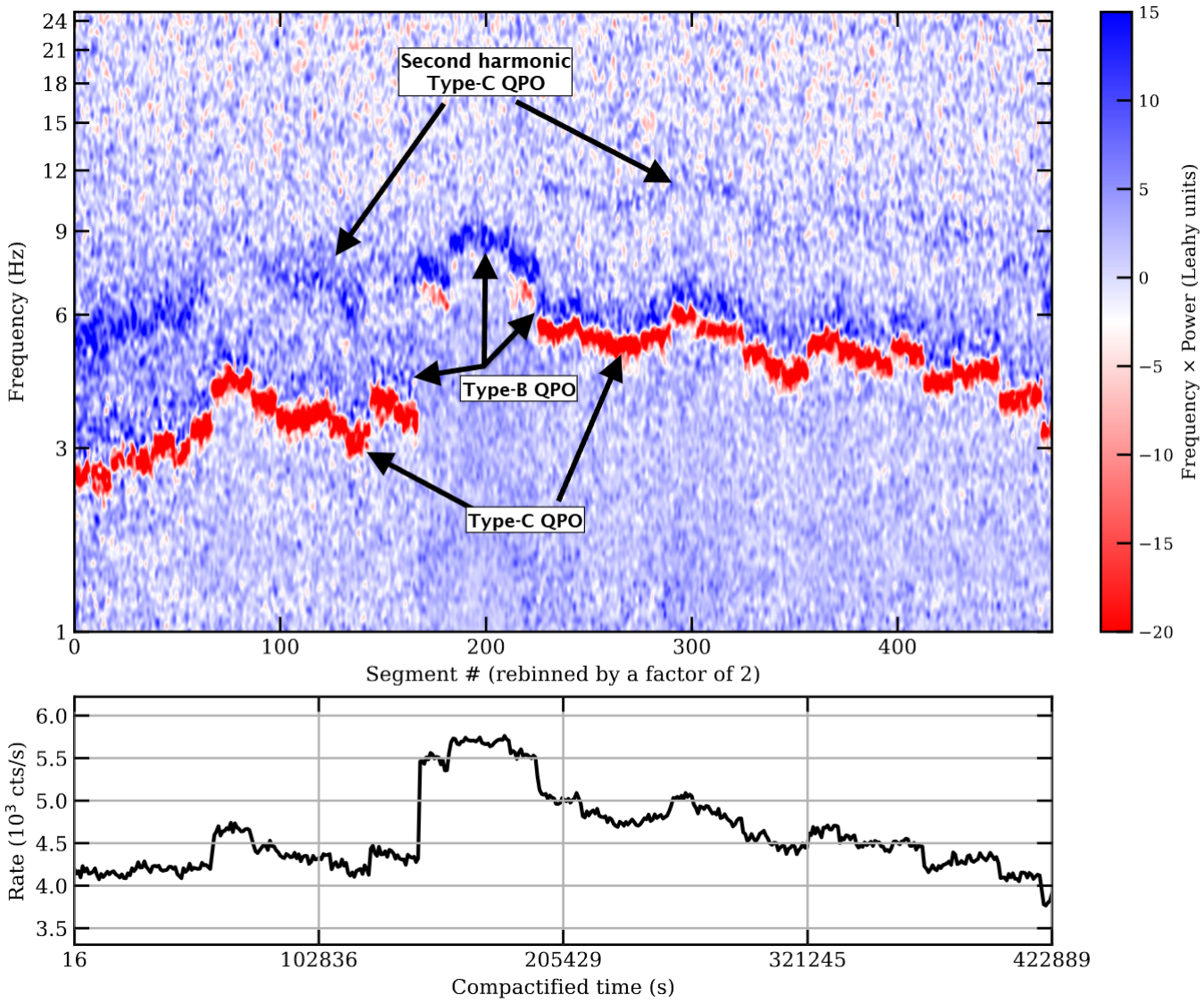}
	\vspace{-5mm}
    \caption{Dynamical imaginary cross-spectrum of Swift J1727.8$-$1613 between MJD 60204 and 60209. The photon count rate of the light curve in the bottom panel is the sum of the LE ($2-10$ keV) and HE ($28-200$ keV) count rates. The dynamical imaginary cross-spectrum is given in Leahy units~\citep{1983ApJ...266..160L}; the segments in the X-axis have a duration of $\sim 60$ seconds. We do not show the time gaps between observations to produce a more compact plot; the time gaps can be seen in Fig.~\ref{fig:flare_light_zoomin}. Compactified time is the true elapsed time in seconds since MJD 60204.22117, with the gaps in the data removed from the plot. The arrows identify the Type-B QPO, and the Type-C QPO and its second harmonic.} 
    \label{fig:dy_imag}
\end{figure}

While we have shown that the Type-B and Type-C QPOs coexist over the time intervals used to compute the PDS and CS (typically $\sim 500-1500$ seconds; see Table~\ref{tab:rep_obs}), it remains to be seen whether this holds true over shorter intervals. Given that the Type-B QPO sometimes disappears for periods of a few hundred seconds~\citep[e.g.;][]{2021MNRAS.505.3823Z, 2023MNRAS.519.1336P}, it is also possible that the QPOs alternate in time rather than being present simultaneously.
\citet{2025ApJ...984L..53W} present the dynamical power spectra (known as the spectrogram in signal processing) of Swift J1727.8$-$1613 around the soft X-ray flare on MJD 60206 and find that, at the same time as the jet knots are ejected, the QPO frequency suddenly increases \citep[see also Figure 2 of][]{2025A&A...699A...9J}. 
However, in this case, the power spectrum alone is not sufficient to separate the two components. Since the two QPOs exhibit very different phase lags (see Fig.~\ref{fig:flare_par}), a more suitable tool to track their evolution separately is the time-frequency plot of the imaginary part of the cross spectrum. In Fig.~\ref{fig:dy_imag} we show the dynamical imaginary part of the CS\footnote{In Fig.~\ref{fig:dy_imag} we show the dynamical imaginary part of the CS. As far as we are aware, there is no established name for this quantity in signal processing. By analogy with the spectrogram, it could be referred to as an imaginary cross-spectrogram.} between the HE ($28-200$ keV) and LE ($2-10$ keV) data. This Figure shows a sharp flip of the lags, from negative for the Type-C QPO (red) to positive for the Type-B QPO (blue).
At a compactified time of $\sim 150$ ks, after a gap in the observations, the source count rate increases from $\sim 4.3 \times 10^3$ counts s$^{-1}$ to $\sim 5.5 \times 10^3$ counts s$^{-1}$. At the same time, the frequency of the Type-C and Type-B QPOs increase from $\sim  3.6$ Hz to $\sim 6.7$ Hz and from $\sim 4.4$ Hz to $\sim 7.5$ Hz, respectively. The second harmonic of the Type C QPO, which was at $\sim 7.2$ Hz before the gap, disappears after the gap. At the peak of the flare, only the Type-B QPO is detected, reaching a maximum frequency of $\sim 9$ Hz. After the flare, at a compactified time of $\sim 190$ ks, when the frequency of the Type-B QPO decreases to $\sim 6.2$ Hz, the Type-C QPO and its second harmonic reappear at $\sim 5.5$ Hz and $\sim 11$ Hz, respectively.
This plot shows that the Type-C and Type-B QPOs coexist simultaneously over time scales as short as $\sim 30-60$ s.

\section{Discussion}
\label{sec:discussion}

Our analysis of a soft X-ray flare in Swift J1727.8$-$1613 on 2023 September 19 (MJD 60206) with \textit{Insight}-HXMT reveals that, at the peak of the flare, the source undergoes a brief but clear transition into the soft-intermediate state (SIMS), marked by the simultaneous appearance of discrete jet ejections, a sharp drop in broadband noise in the 2$-$10 keV energy band, and a narrow, coherent QPO with a characteristic ``U''-shaped phase-lag spectrum and a quality factor $Q \geq 6$, which are hallmarks that strongly support its identification as a Type-B QPO. Crucially, this transition is accompanied by the complete disappearance of the Type-C QPO, which had been persistently present before the flare and re-emerges only after it. Yet, perhaps most notably, the Type-B QPO is not confined to this transient excursion into the SIMS: we detect, for the first time, the Type-B QPO consistently across all our observations in the hard-intermediate state (HIMS), where it appears as a broad shoulder of the Type-C QPO in the PDS.

Thanks to the continuous and well-sampled coverage provided by \textit{Insight}-HXMT, we trace for the first time a smooth and coherent evolution of both the Type-C and Type-B QPOs across the flare. These two components remain clearly distinguishable throughout, primarily by their markedly different phase lags. We find that the properties of these QPOs are closely linked to the hardness ratio, both during the rise phase of the LE light curve when the hardness ratio decreases, and during the decay phase when the hardness ratio increases.
As the hardness ratio decreases: (i) the frequencies of both QPOs increase, with the Type-B QPO consistently maintaining a slightly higher frequency than the Type-C; (ii) the fractional rms amplitude of the Type-C QPO drops steadily in both the LE and HE bands, vanishing altogether at the flare's peak; (iii) the Type-B QPO becomes increasingly prominent in the HE band, peaking in amplitude at the same time the Type-C QPO disappears, while its amplitude in the LE band shows a mild decline; (iv) the Type-B QPO becomes progressively more coherent, reaching $Q \geq 6$ during the flare, whereas the Type-C QPO displays a roughly constant coherence throughout the observations we used, except for the observation at the peak of the flare where the Type-C QPO is not  detected.

Our results indicate that the Type-B QPO appears in all our observations considered here, including those before and after the flare, when a Type-C QPO is present in the PDS and the source is in the HIMS~\citep[e.g.;][]{2024MNRAS.529.4624Y, 2025A&A...697A.229R}. These results provide new insight into the physical and geometrical origins of the QPOs. We will discuss and offer an interpretation of these findings in the following subsections.

\vspace{-4mm}
\subsection{Type-B QPOs and SIMS}
\label{sec:Type-B QPOs and SIMS}

As a black-hole transient transitions from the HIMS into the SIMS, three key properties are typically observed~\citep[e.g.;][]{2011BASI...39..409B}. The first is the appearance of the Type-B QPO. One of the main properties of the Type-B QPO, as summarized by \citet{2005ApJ...629..403C}, is its narrow profile, with $Q \geq 6$. Over the past two decades, the Type-B QPOs observed in BHXBs have consistently exhibited this property~\citep[e.g.;][]{2011MNRAS.418.2292M, 2021MNRAS.505.3823Z}. The second property is a decrease of the fractional rms amplitude of the broadband noise, typically falling below 10\% in the $2-15$~keV band in the SIMS~\citep{2011MNRAS.410..679M}, compared to $\geq 30$\% in the LHS and 10$-$30\% in the HIMS. The third indicator of the source transitioning into the SIMS is the appearance of discrete radio jet ejections~\citep{2020ApJ...891L..29H, 2020NatAs...4..697B, 2021MNRAS.505.3393W}.
Contrary to the steady jet observed in the LHS and HIMS~\citep[e.g.;][]{2022NatAs...6..577M, 2024ApJ...971L...9W, 2025ApJ...984L..53W}, these discrete ejections are optically thin.

In the case of Swift~J1727.8$-$1613, during the X-ray flare (MJD 60206), marked with the gray region in Fig.~\ref{fig:flare_light_zoomin},~\ref{fig:flare_par}, and~\ref{fig:flare_rms}, the Type-B QPO is also narrow, with $Q \sim 6.4-13.7$  ( $1/Q \sim 0.16-0.07$), consistent with the $Q$ values of other Type-B QPOs detected in the SIMS in other sources~\citep[e.g.;][]{2011MNRAS.418.2292M, 2021MNRAS.505.3823Z}. 
This QPO exhibits a ``U''-shaped phase lag spectrum, with its minimum occurring at approximately $3$~keV, except the observation at the peak of the flare, where the minimum shifts to around $9$~keV. The overall trend of the phase lag spectrum is consistent with those observed in the Type-B QPOs from other sources, such as MAXI~J1348$-$630~\citep{2020MNRAS.496.4366B, 2021MNRAS.501.3173G}, MAXI~J1820$+$070 \citep{2023MNRAS.525..854M}, GX~339$-$4~\citep{2023MNRAS.519.1336P}, MAXI~J1535$-$571~\citep{2023MNRAS.520.5144Z} and Swift~J1728.9$-$3613~\citep{2024RAA....24c5001K}. In the cases of MAXI~J1348$-$630~\citep{2020MNRAS.496.4366B, 2021MNRAS.501.3173G}, MAXI~J1820$+$070 \citep{2023MNRAS.525..854M} and GX~339$-$4~\citep{2023MNRAS.519.1336P}, the minimum phase lag of the Type-B QPOs is at $\sim 2-3$~keV. In contrast, the Type-B QPO observed in MAXI J1535$-$571 shows a phase lag spectrum with a minimum at $\sim 6-8$~keV~\citep{2023MNRAS.520.5144Z}, although the minimum could lie at even higher energies, as the NICER energy range is not sufficient to clearly observe the upturn. On the other hand, in Swift~J1728.9$-$3613~\citep{2024RAA....24c5001K}, a 5.76-Hz Type-B QPO has a minimum phase lag at around $2-3.5$~keV, while a 5.46-Hz Type-B QPO displays a nearly flat phase lag spectrum above $\sim4$~keV within the NICER energy band.

Simultaneously, the total ($0.01-50$ Hz) fractional rms in the LE $2-10$ keV band drops below 9\% during the soft X-ray flare (Fig.~\ref{fig:flare_rms}), which is consistent with the typical rms (below 10\%) observed in the SIMS~\citep{2005Ap&SS.300..107H, 2011MNRAS.410..679M}. In addition, during the soft X-ray flare on 2023 September 19 (MJD 60206), three jet knots are ejected and identified in the VLBA image of Swift~J1727.8$-$1613~\citep{2025ApJ...984L..53W}. Therefore, from all this, we conclude that during the flare on MJD 60206, Swift~J1727.8$-$1613 enters the traditional SIMS.

However, we also detect Type-B QPOs throughout all the observations, from MJD 60204 to 60209, that are taken from the HIMS and this brief transition into the SIMS. 
This result raises questions about the traditional paradigm that associates Type-B QPOs exclusively with the SIMS.
In Swift~J1727.8$-$1613, the Type-B QPO is present both before and after the soft X-ray flare. At the same time, the Type-B QPO has a broad profile and appears as a shoulder of the Type-C QPO, resembling the shoulder of the Type-C QPOs observed in the past~\citep{1997A&A...322..857B, 2002ApJ...572..392B, 2020MNRAS.496.5262V, 2024MNRAS.527.9405M}. As shown in Fig.~\ref{fig:flare_representative_observations_spectra}, the fractional rms and phase lag spectra of this shoulder of the Type-C QPO are consistent with those of the QPO that we identify as Type-B QPO in the flare. We therefore classify this shoulder of the Type-C QPO as a Type-B QPO.

In Section~\ref{sec:Investigate Type-B QPO signature in other observations}, we investigate two observations before those we presented above and find that the shoulder of the Type-C QPO still appears in the these two observations with similar properties as those of the Type-B QPO during the X-ray flare. These indicate that, at least in Swift~J1727.8$-$1613, the Type-B QPO does not suddenly appear in the SIMS, but it is already present in the HIMS.

\citet{2012MNRAS.427..595M} already detected simultaneous Type-C and -B QPOs in GRO~J1655$-$40~\citep[see also;][]{2023MNRAS.525..221R}, although such occurrences are very rare. In that case, the source was in the so-called ultra-luminous (or anomalous) state~\citep[see Figure 1 of][]{2012MNRAS.427..595M}. In the case of Swift~J1727.8$-$1613, however, we detect both QPOs simultaneously during the HIMS, marking the first time such a phenomenon is observed in this state.

In previous studies, transitions between the Type-C and Type-B QPOs almost always happened during gaps in the X-ray coverage. Transitions were caught successfully in the cases of MAXI~J1820$+$070~\citep{2020ApJ...891L..29H, 2023MNRAS.525..854M} and MAXI~J1348$-$630~\citep{2022ApJ...938..108L}. In both cases the Type-C QPO suddenly switched to the Type-B QPO, from an 8-Hz Type-C QPO to a 4.5-Hz Type-B QPO in MAXI~J1820$+$070 and from a 9.2-Hz Type-C QPO to a 4.8-Hz Type-B QPO in MAXI~J1348$-$630. 
A more recent study by \citet{2025MNRAS.538.1143L} suggests that, in MAXI~J1820$+$070, the Type-B QPO also appears before the source transitions into the SIMS. However, they do not analyze the phase-lag spectrum or explicitly identify the QPO types.
It remains to be seen whether those analyses missed weaker forms of the Type-C and Type-B QPOs because they relied solely on the PDS. In the case of Swift~J1727.8$-$1613, indeed the strength of the Type-C QPO decreases and the QPO disappears at the peak of the soft X-ray flare on MJD 60206, at the same time when the strength of the Type-B QPO reaches the maximum (Fig.~\ref{fig:flare_rms} and~\ref{fig:flare_hfd}). However, due to the close proximity of the frequencies of the Type-C and Type-B QPOs in Swift~J1727.8$-$1613, the switch between the Type-C and Type-B QPOs is not distinctly apparent in the PDS. In fact, the Type-B QPO observed in the flare was interpreted as a Type-C QPO in previous studies~\citep{2024ApJ...968..106Z, 2025ApJ...984L..53W, 2025MNRAS.tmp..712B, 2025arXiv250618857M, 2025arXiv250416391D}. 
Compared to the traditional method that relies solely on PDS analysis to detect timing variability components, the joint fit of PDS and CS \citep{2024MNRAS.527.9405M} makes better use of the available data and can reveal variability components that are otherwise hidden when using the PDS alone~\citep[e.g.;][]{2025A&A...696A.128B, 2025A&A...696A.237F, 2025arXiv250507938B, 2025ApJ...990...43R}. 
Taking advantage of this approach, in this paper we discover a Type-B QPO in Swift~J1727.8$-$1613 in the HIMS, which can be distinguished from the Type-C QPO when both QPOs are present simultaneously.

This result also supports the interpretation that the profile of the Type-C QPO and its shoulder, here identified as the Type-B QPO, consists of two distinct components, which, as we have done here, are typically fitted with two separate Lorentzian functions: At the peak of the soft flare, the shoulder (that is, the Type-B QPO) is observed in isolation, while the main QPO, the Type-C QPO, is absent.
\citet{2024MNRAS.527.9405M} identified the shoulder of the Type-C QPO in GRS~1915$+$105 and GX~339$-$4 using the model of two separate Lorentzians. We speculate that (some of) those shoulders could correspond to the Type-B QPO.
Our findings here challenges interpretations in which the shoulder is not a separate oscillation but rather part of a single, asymmetric, profile caused by, for instance, multiplicative variability components in the time domain~\citep[e.g.;][]{2013MNRAS.434.1476I}. It remains to be seen whether the idea that the QPOs are due to multiplicative signals in the time domain can be extended to the real and imaginary parts of the CS, and whether such a model can accurately predict the lags and coherence function, as is the case with our assumption of additive components.

\subsection{Tracing accretion-flow geometry through the Type-B and Type-C QPOs}

Profiting from the good coverage in both X-ray and radio, \citet{2020ApJ...891L..29H} found evidence that in MAXI~J1820$+$070 the appearance of the Type-B QPO is connected with the discrete jet ejection events.
In the case of Swift~J1727.8$-$1613, \citet{2025ApJ...984L..53W} detected that jet knots were ejected at the time of the soft X-ray flare, on 2023 September 19, when we detect the Type-B QPOs, which would point to a similar connection. However, as we explain in Section~\ref{sec:Type-B QPOs and SIMS}, in Swift~J1727.8$-$1613 Type-B QPOs are detected throughout all observations from MJD 60204 to 60209, not exclusively during the brief transition of the source to  the SIMS. The fact that Type-B QPOs can exist without accompanying discrete jet ejections suggests that the Type-B QPOs are not caused by, or directly correlated to, those ejections, as was previously suggested by \citet{2024MNRAS.533.4188C}.

\citet{2016MNRAS.460.2796S} performed phase-resolved spectroscopy of the Type-B QPO in GX 339$-$4 and suggested that this QPO can be explained by a precessing jet-base corona~\citep[see also;][]{2020A&A...640L..16K}. Using phase-resolved spectroscopy of the Type-C QPOs, \citet{2015MNRAS.446.3516I, 2016MNRAS.461.1967I, 2017MNRAS.464.2979I} suggested that this type of QPOs can be explained by a precessing hot inner accretion flow, located inside the truncation radius of the accretion disc. Previous studies of their fractional rms and phase lag spectra using the vKompth model \citep{2020MNRAS.492.1399K, 2022MNRAS.515.2099B}, which introduces a feedback factor representing the fraction of the photons Comptonized in the corona that return to the seed-photon source, suggested that the Type‐C QPOs are associated with a disc corona, located within, or partially covering the accretion disc~\citep{2022NatAs...6..577M, 2022MNRAS.513.4196G, 2023MNRAS.525..854M, 2023MNRAS.525..221R, 2023MNRAS.520..113R, 2025ApJ...980..251A}, whereas Type‐B QPOs are linked to a jet-base corona~\citep{2023MNRAS.519.1336P, 2023MNRAS.520.5144Z, 2023MNRAS.525..854M, 2025ApJ...980..251A}. 

In Swift J1727.8$-$1613 the frequencies of both the Type-C and Type-B QPOs are anti-correlated with the hardness ratio (Fig.~\ref{fig:flare_hfd}). 
In the case of Swift J1727.8$-$1613, \citet{2025arXiv250618857M} find that the Type-C QPO tracks the truncation radius of the accretion disc. 
Therefore, the observed trend for the Type-C QPO naturally follows: as the accretion disc moves inward and the disc-corona that produces the Type-C QPO contracts, the increase of the disc emission leads to a softer energy spectrum.
In Swift J1727.8$-$1613 the Type-B QPO frequency is slightly higher than the frequency of the Type-C QPO throughout the observations. Under the assumption that higher frequency corresponds to a smaller emission region, this would indicate that the region responsible for the Type-B QPO is smaller than the disc–corona region responsible for the Type-C QPO.  Both the Poynting-Robertson cosmic battery~\citep[][]{2012A&A...538A...5K} and the magnetically arrested disc~\citep{2016MNRAS.462..636A, 2022MNRAS.511.3795N} models propose that a strongly magnetic field region, serving as the jet base, forms naturally at the innermost region of the disc-corona. 
It is therefore possible that the frequency of the Type-B QPO is linked to the characteristic radius of this strong magnetic field region. Our results would then suggest that, as the truncation radius of the accretion disc moves inward (the hardness ratio decreases), the characteristic radius of both the disc-corona (responsible for the Type-C QPO) and the jet-base corona (responsible for the Type-B QPO) decrease.

We also found that, as the hardness ratio decreases, the fractional rms amplitude of the Type-C QPO both in the LE and HE energy bands decreases, and eventually the QPO is no longer detected (Fig.~\ref{fig:flare_hfd}). This may indicate that the disc-corona (responsible for the Type-C QPO) shrinks and then becomes very weak, or even disappears, as the truncation radius of the accretion disc moves inward (the hardness ratio decreases). At the same time,  the jet-base corona (responsible for the Type-B QPO) appears to dissipate more power, as suggested by the fact that the fractional rms amplitude of the Type-B QPO in the HE energy band increases as the hardness decreases (Fig.~\ref{fig:flare_hfd}). However, the fractional rms amplitude of the Type-B QPO in the LE energy band (Fig.~\ref{fig:flare_hfd}) decreases as the hardness decrease.
Under the assumption that the Type-B QPO originates from a coupling between the accretion disc and the jet-base corona, this behavior can be explained by an increase in the Lorentz factor of the electrons in the jet-base corona as the source transitions from the HIMS to the SIMS~\citep{2004MNRAS.355.1105F}. This results in fewer high-energy photons from the jet-base corona returning to the disc~\citep{2021NatCo..12.1025Y}. As a consequence, the feedback is reduced, weakening the coupling between the disc and the jet-base corona.

\citet[][]{2025ApJ...984L..53W} detected jet ejections on 2023 September 19 (MJD 60206), that occur simultaneously with a soft X-ray flare. At the beginning of the flare the frequencies of both the Type-C and Type-B QPOs show a rapid increase, as shown in Fig.~\ref{fig:flare_par}. Some theoretical models predict that discrete jet ejections occur when the magnetic pressure becomes dominant as the magnetic field accumulates in the jet-base corona~\citep{2012A&A...538A...5K, 2022ApJ...924L..32R, 2023PhRvR...5d3023Z}. In these scenarios, the built-up magnetic field in the inner regions of the accretion flow can become unstable, triggering the ejection of a substantial amount of accreting material and allowing the accumulated magnetic field to escape.  If this interpretation holds, the soft X-ray flare in Swift J1727.8$-$1613 shown in Fig.~\ref{fig:flare_light_zoomin} could be explained naturally by the rapid inward motion of the accretion disc caused by the (local) disruption of the magnetic pressure resulting from the escape of the magnetic field. This scenario also offers a natural explanation for the observed rapid increase in the frequencies of both the Type-C and Type-B QPOs at the beginning of the flare, possibly due to a rapid and simultaneous shrinkage of the disc-corona and jet-base corona regions when the accretion disc moves inward rapidly. 
The disappearance of the Type-C QPO at the peak of the flare suggests that this unstable process destroys largely or even completely the disc-corona that produces the Type-C QPO.
The discovery of a rapid inward movement of the disc inner radius during the flare in Swift J1727.8$-$1613~\citep{2025arXiv250618857M} further supports a change in the geometry of the accretion flow~\citep[see also][]{2025arXiv250801384H, 2025ApJ...993...40X}.
In this scenario, in the decay phase of the flare, the magnetic field  would accumulate again,  the sizes of the corona regions would increase again, and the frequency of the QPOs would  decrease again.

Additionally, our observations show that the Type-B QPO transitions from a broad to a narrow profile, evolving into a typical Type-B QPO feature during the flare shown in Fig.~\ref{fig:flare_par}. This transition occurs as the source approaches the peak of the LE light curve, coinciding with an increase of the Type-B QPO frequency. Thus, for the Type-B QPO, a higher quality factor is associated with a higher frequency. We suggest that the quality factor of the Type-B QPO in Swift J1727.8$-$1613 may be linked to the characteristic width of the jet base corona. The same idea has been also proposed by  \citet{2020A&A...640L..16K}. They suggested that the absence of Type-B QPOs in the LHS and HIMS may be due to by a wide jet-base corona, although the jet-base corona may also be precessing.
As we mentioned before, the frequency of the Type-B QPO is consistently higher than that of the Type-C QPO in Swift J1727.8$-$1613, in contrast to the cases of MAXI~J1820$+$070~\citep{2020ApJ...891L..29H, 2023MNRAS.525..854M} and MAXI~J1348$-$630~\citep{2022ApJ...938..108L}. This discrepancy suggests that the location of the jet-base corona may differ between systems. 
In this scenario, in the case of Swift J1727.8$-$1613, the jet-base corona would be situated near the disc plane very close to the black hole. In contrast, for systems where the Type-B QPO has a lower frequency than the Type-C QPO during the transition from the HIMS to the SIMS, the jet-base corona would be farther away from both the disc plane and the black hole.

\section{Conclusions}
\label{sec:summary}

In this paper we present results that challenge the traditional paradigm that associates Type-B QPOs exclusively with the SIMS and discrete jet ejections, and instead suggest a more nuanced picture in which this QPO component emerges under a wider range of accretion conditions. Our conclusions are as follows:
\begin{enumerate}[.]
  \item The source Swift J1727.8$-$1613 undergoes a brief but clear transition into the SIMS during the soft X-ray flare on MJD 60206.
  \item The Type-B QPO is detected during both the HIMS and the SIMS, rather than being limited only to the SIMS.
  \item In the HIMS, the Type-B QPO co-exists with the Type-C QPO, whereas in the transition to the SIMS the Type-C QPO weakens and eventually disappears when the source is in the SIMS.
  \item The frequency of the Type-B QPO of Swift J1727.8$-$1613 is consistently higher than that of the Type-C QPO and is anti-correlated with the hardness ratio.
  \item The distinct evolutionary trends of the Type-B and Type-C QPOs suggest that the Type-C QPO is associated with a disc-corona, while the Type-B QPO is linked to a jet-base corona.
\end{enumerate}

The joint-fitting technique of the PDS and CS proposed by~\citet{2024MNRAS.527.9405M} has demonstrated significant potential for uncovering QPOs that are otherwise hidden in the PDS alone~\citep[e.g.;][]{2025A&A...696A.128B, 2025A&A...696A.237F, 2025arXiv250507938B}. Applying this technique to archival data of other sources may allow the detection of weak Type-B QPOs that were previously missed using conventional PDS analysis. Such re-investigations could provide new insights into the occurrence and properties of Type-B QPOs, particularly during phases when they are too faint to dominate the PDS.

\begin{acknowledgements}
We thank the referee for the insightful suggestions that helped improve the clarity of our work. MM acknowledges the research programme Athena with project number 184.034.002, which is (partly) financed by the Dutch Research Council (NWO). PJ acknowledges support from the China Scholarship Council (CSC 202304910058). FG acknowledges support by PIBAA 1275 and PIP 0113 (CONICET). FG was also supported by grant PID2022-136828NB-C42 funded by the Spanish MCIN/AEI/ 10.13039/501100011033 and “ERDF A way of making Europe”.
F.M.V. is supported by the European Union's Horizon Europe research and innovation programme through the Marie Sklodowska-Curie grant agreement No. 101149685.
\end{acknowledgements}

\vspace{-4mm}

\bibliographystyle{jwaabib}
\bibliography{references}

\appendix

\section{Rotated CS}
\label{sec:rotated cs}

\begin{figure}
    \vspace{-20mm}
	\includegraphics[width=\columnwidth]{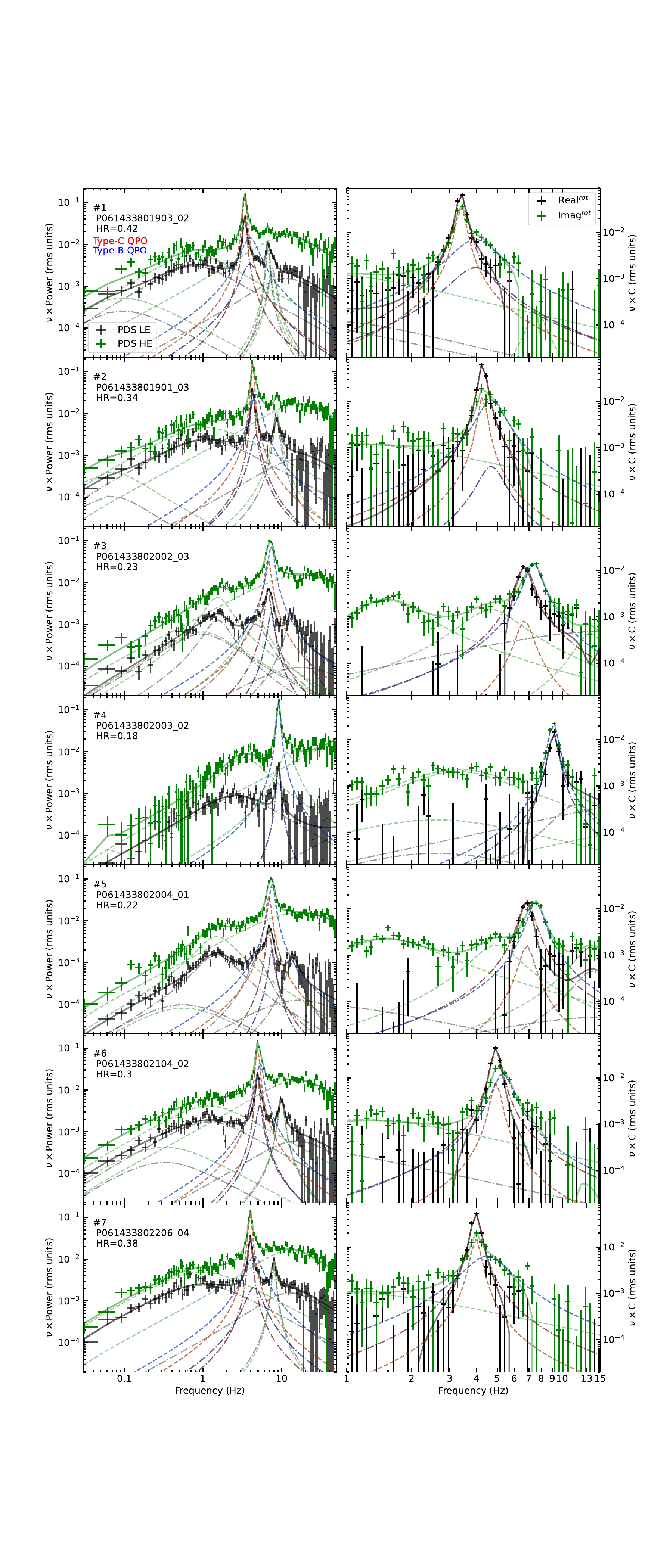}
	\vspace{-28mm}
    \caption{Same as Fig~\ref{fig:flare_representative_observations}, but the cross vectors are rotated counterclockwise by $\pi/4$ rad. The PDS and CS are rebinned in frequency by a factor $\approx 1.047 = 10^{1/50}$.} 
    \label{fig:flare_representative_observations_1}
\end{figure}

In this section, we rotate the cross vectors counterclockwise by $\pi/4$ rad~\citep{2024MNRAS.527.9405M}, which is equivalent to rotating the coordinate axes clockwise by $\pi/4$ rad. In the right panel of Fig.~\ref{fig:flare_representative_observations_1}, we show the rotated real and imaginary parts of the cross vector and highlight the contributions from the Type-C (red) and Type-B (blue) QPOs.
The Type-C QPO almost always dominates the rotated real part, except in Obs. \#5 at the flare peak, where the Type-C QPO disappears. The Type-B QPO dominates the rotated imaginary part in Obs. \#3, \#4, \#5, and \#6. Notably, in Obs. \#3 and \#5 (panels 3 and 5 in Fig.~\ref{fig:flare_representative_observations_1}), the peaks of the rotated real and imaginary parts do not coincide: the peak of the Type-C QPO coincides with that of the rotated real part, while the peak of the Type-B QPO coincides with that of the rotated imaginary part.
This behavior further supports our conclusion that the Type-C and Type-B QPOs represent independent variability components.

\begin{figure}
    \vspace{-5mm}
	\includegraphics[width=\columnwidth]{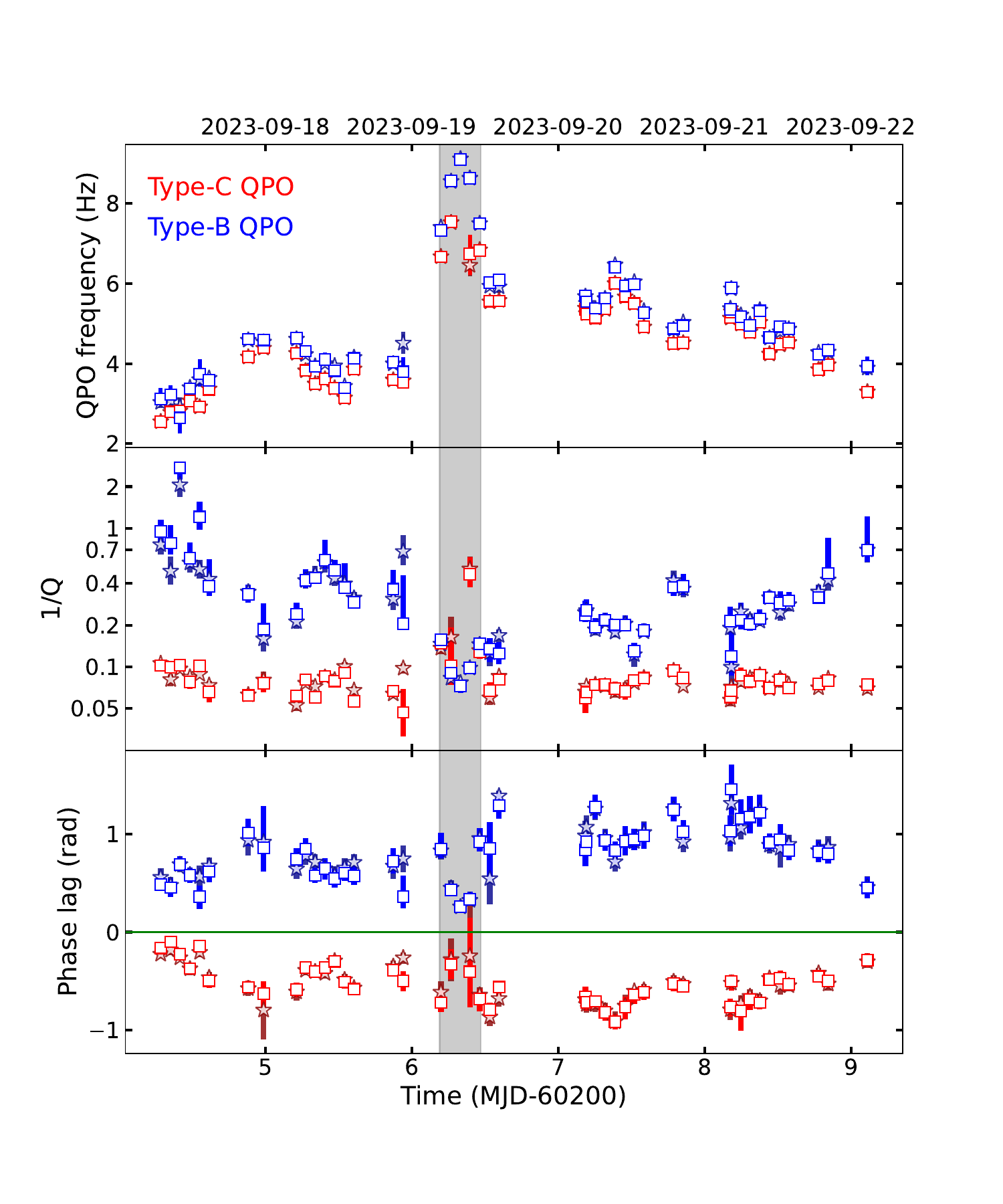}
	\vspace{-10mm}
    \caption{Same as Fig.~\ref{fig:flare_par}, but obtained from the fit to the rotated CS~\citep{2024MNRAS.527.9405M}. The squares represent the parameters from the rotated CS, whereas the stars denote the same parameters from the non-rotated CS.}
    \label{fig:flare_par_rot}
\end{figure}

In Fig.~\ref{fig:flare_par_rot}, we present the QPO parameters obtained from the fits using the rotated CS~\citep{2024MNRAS.527.9405M}. As expected, the parameters of the two QPOs are consistent within the 1$\sigma$ uncertainties with those obtained from the fits in Fig.~\ref{fig:flare_par}. This indicates that rotating the cross vector does not affect the fit results.

\section{Coherence function and phase lags}
\label{sec:coh}

\begin{figure}
	\includegraphics[width=\columnwidth]{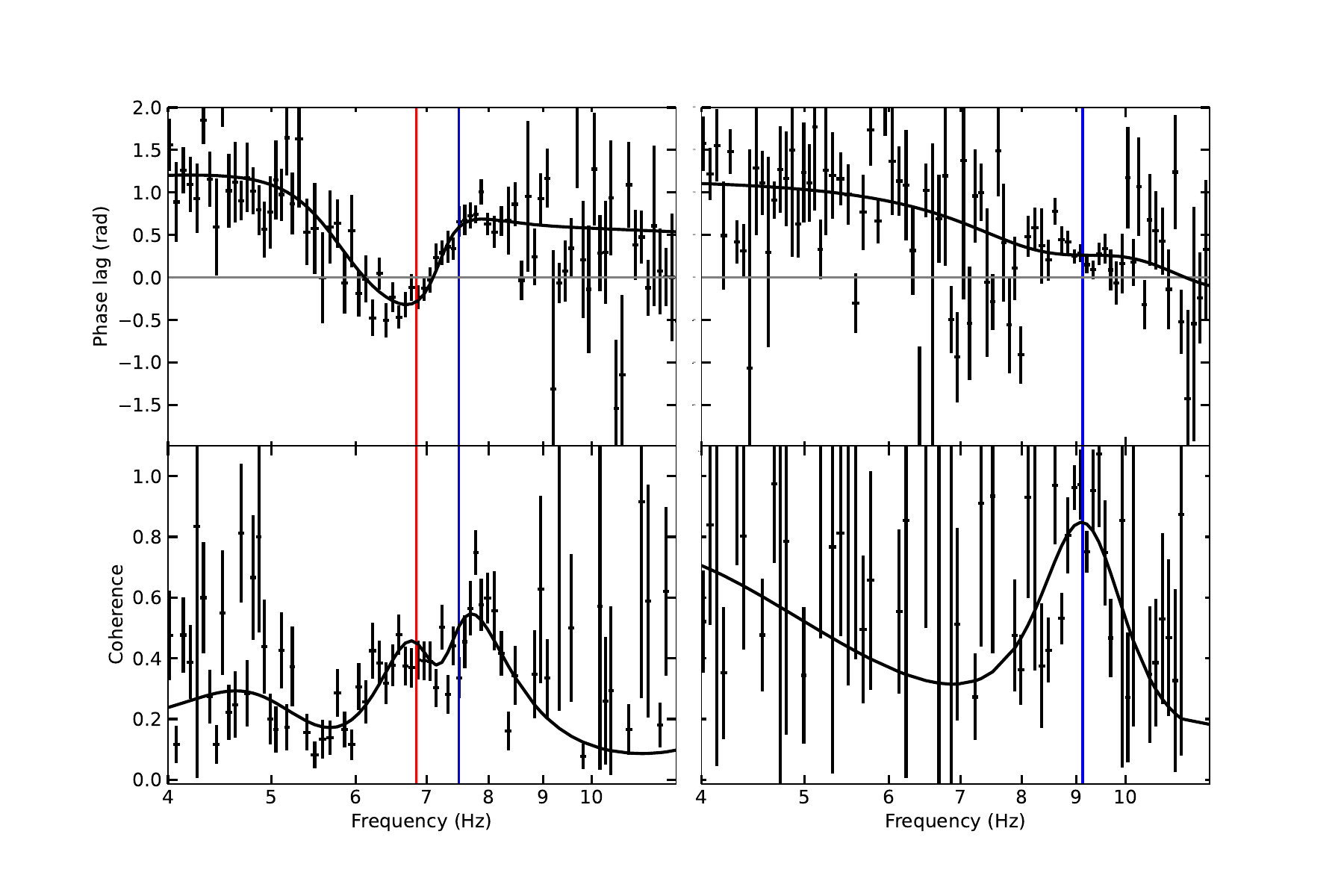}
	\vspace{-5mm}
    \caption{Phase lags and the coherence function of Obs \#5 (left panel) and \#4 (right panel) in Table~\ref{tab:rep_obs}. The red vertical line indicates the centroid frequency of the Type-C QPO, while the blue vertical line indicates that of the Type-B QPO. The models are not fitted to the data, but predicted on the basis of the parameters of the Lorentzians fitted to the PDS and the real and imaginary parts of the CS.}  
    \label{fig:coh}
\end{figure}

In Fig.~\ref{fig:coh} we plot the phase lags and the coherence function of Obs \#5 (left panel) and \#4 (right panel) in Table~\ref{tab:rep_obs}. In Obs \#5, where both the Type-C and Type-B QPOs are significantly detected (Fig.~\ref{fig:flare_representative_observations}), the coherence peaks at the centroid frequencies of both the Type-C and the Type-B QPOs.  The coherence function reaches a values of $\sim$0.4 at the maximum, with a dip in between, where the profiles of the two QPOs overlap, because of the decrease of the coherence where the two signals overlap. At the peak of the flare (Obs \#4), where only one QPO is present in the PDS and CS, the coherence shows a single narrow peak at the centroid frequency of the Type-B QPO that reaches a maximum value of  $\sim$0.8. 
The higher value of the coherence in this case is consistent with the fact that, contrary to Obs \#5, in Obs \#4 there is only one QPO, which is then not affected by the loss of the coherence produced by the other QPO in \#5. The small drop from unit coherence in this case is due to other variability components in the data that overlap in frequency with the Type-B QPO (see Fig.~\ref{fig:flare_representative_observations}).
The phase lags of Obs \#5 exhibit a sharp dip with a negative minimum at the Type-C QPO frequency and a plateau at the frequency of the Type-B QPO, whereas in Obs \#4 the phase lags remain positive near the QPO frequency, owing to the disappearance of the Type-C QPO.
We note that in both panels in Fig.~\ref{fig:coh} the continuous lines were not fitted to the lags and the coherence function, but are the prediction of the best fitting model to the PDS and CS.
As explained in \citet{2024MNRAS.527.9405M}, this consistency provides further support for our conclusion that the Type-B QPO, appearing as the shoulder of the Type-C QPO, is an independent variability component. 

\section{Type-B QPO signature in other observations}
\label{sec:Investigate Type-B QPO signature in other observations}

\begin{figure}
	\includegraphics[width=\columnwidth]{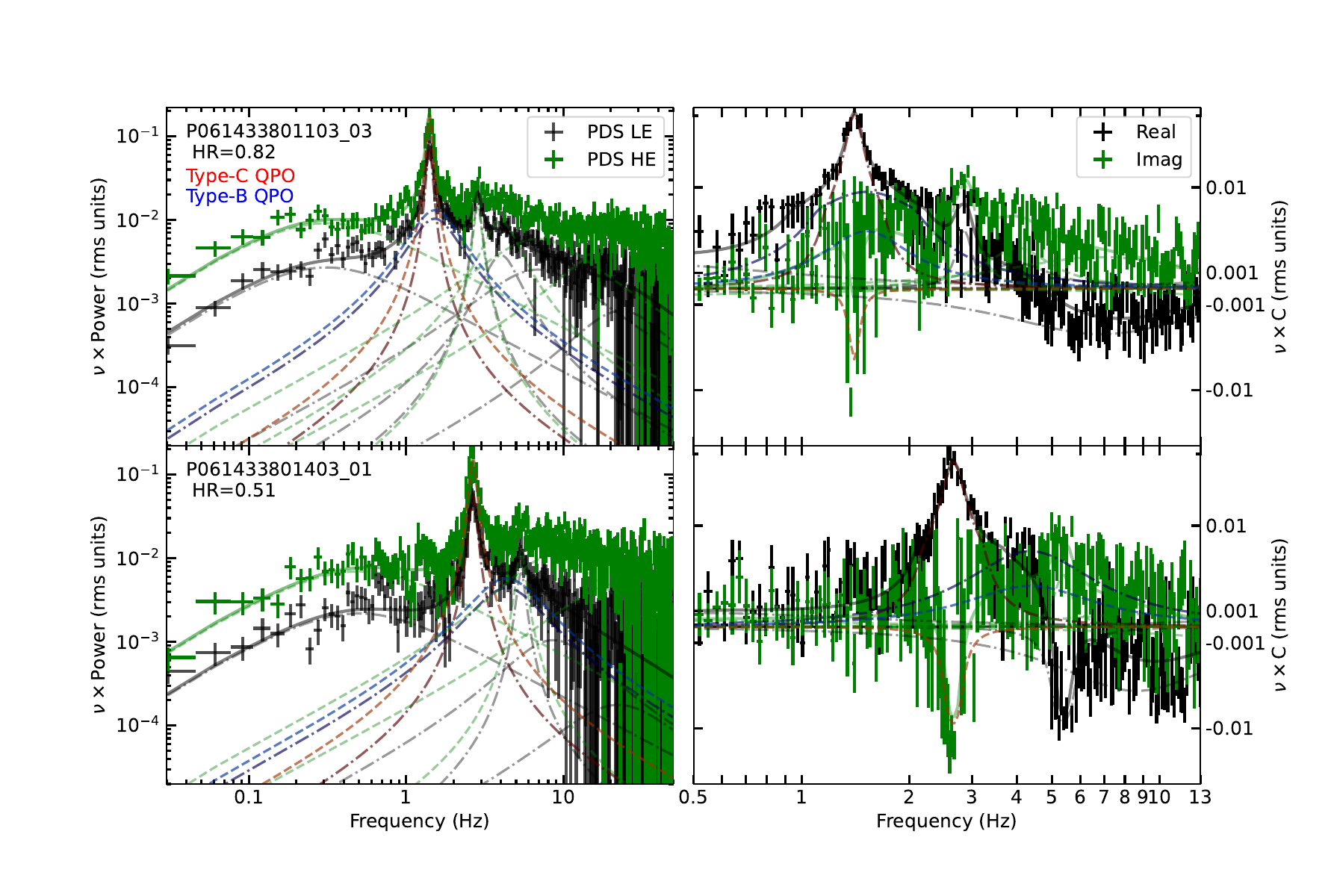}
	\vspace{-5mm}
    \caption{Same as Fig.~\ref{fig:flare_representative_observations}, but for observations P061433801103\_03 (left panel) and P061433801403\_01 (right panel).}  
    \label{fig:rep_before_flare}
\end{figure}

To investigate whether the Type-B QPO is consistently present in other observations in the HIMS, we analyze two observations before the ones presented above. We fit jointly the LE $2-10$~keV PDS and HE $28-200$~keV PDS, as well as the real and imaginary parts of the corresponding CS between HE $28-200$~keV and LE $2-10$~keV data, as we do in Section~\ref{sec:Joint-fit of power and cross spectra}. In Fig.~\ref{fig:rep_before_flare} we plot the PDS, along with the real and imaginary parts of the CS, for these two observations, in the same format as presented in Fig.~\ref{fig:flare_representative_observations}.
The first observation, P061433801103\_03, taken on MJD 60197.45, shows a Type-C QPO with a centroid frequency of $1.41 \pm 0.01$ Hz and a phase lag of $-0.04 \pm 0.04$ rad between the HE $28-200$~keV and the LE $2-10$~keV data. A shoulder of the Type-C QPO has a centroid frequency of $1.42 \pm 0.04$ Hz, a quality factor of $\sim 1.3$, and a phase lag of $0.39 \pm 0.09$ rad. The phase lag of this shoulder is consistent with the values of the Type-B QPOs in Fig.~\ref{fig:flare_par}, and is $4.4\sigma$ different from that of the Type-C QPO. The second observation, P061433801403\_01, taken on MJD 60200.36, exhibits a Type-C QPO with a centroid frequency of $2.65 \pm 0.01$ Hz and a phase lag of $-0.11 \pm 0.04$ rad. A shoulder of the Type-C QPO is also detected in this observation, with a centroid frequency of $3.7 \pm 0.3$ Hz, a quality factor of $\sim0.8$ and a phase lag of $0.48 \pm 0.18$ rad. This phase lag also matches the values of the Type-B QPOs in Fig.~\ref{fig:flare_par}, and differs from that of the Type-C QPO by $3.2\sigma$.
These results show that the shoulder of the Type-C QPO is always present, but outside the soft X-ray flare this shoulder has a broad profile. All this indicates that this shoulder is the same component we identify as the Type-B QPO around the peak of the soft X-ray flare on MJD 60206.

\end{document}